\documentclass[aps,pra,preprint,onecolumn,showpacs,preprintnumbers,amsmath,amssymb,floatfix]{revtex4}
\usepackage{amsthm,graphicx}

\newcommand{\nc}{N_{cr}}
\newcommand{\ncgp}{N^{GP}_{cr}}
\newcommand{\om}{\omega}

\newcommand{\s}{\sigma}
\newcommand{\ta}{\tau}
\newcommand{\so}{\sigma_0}
\newcommand{\se}{\sigma_1}
\newcommand{\si}{\sigma_i}
\newcommand{\la}{\lambda}
\newcommand{\lo}{\lambda_0}
\newcommand{\La}{\lambda}

\newcommand{\e}{\epsilon}

\newcommand{\tv}{|\tau|}
\newcommand{\psir}{|\Psi\rangle}
\newcommand{\psil}{|\Psi^L\rangle}
\newcommand{\psili}{|\Psi^L_i\rangle}
\newcommand{\psile}{|\Psi^L_{i=1}\rangle}
\newcommand{\phir}{|\Phi\rangle}
\newcommand{\phirl}{|\Phi^L\rangle}

\def\r {{\bf r}}

\def\a {\alpha}
\def\b {\beta}

\def\re {\rho_1}
\def\tno {|\tau|}

\begin{document}

\title{Fragmented Many-Body states of definite angular momentum and stability of attractive 3D Condensates}

\author{Marios Tsatsos}\email[Corresponding author, Electronic address:~]{marios.tsatsos@pci.uni-heidelberg.de}
\author{Alexej I. Streltsov}\email{alexej.streltsov@pci.uni-heidelberg.de}
\author{Ofir E. Alon}\email[Present address: Department of Physics, University of Haifa at Oranim, Tivon 36006, Israel]{}
\author{Lorenz S. Cederbaum}\email{lorenz.cederbaum@pci.uni-heidelberg.de}

\affiliation{Theoretische Chemie, Physikalisch-Chemisches Institut, Universit\"at Heidelberg, Im Neuenheimer Feld 229, D-69120 Heidelberg, Germany}

\date{\today}

\begin{abstract}
A three dimensional attractive Bose-Einstein Condensate (BEC) is expected to collapse, when the number of the particles $N$ in the ground state or the interaction strength $\lo$ exceeds a critical value. We study systems of different particle numbers and interaction strength and find that even if the overall ground state is collapsed there is a plethora of fragmented excited states that are still in the metastable region. Utilizing the \emph{configuration interaction} expansion we determine the spectrum of the ground (`yrast') and excited many-body states with definite total angular momentum quantum numbers $0\leqslant L\leqslant N$ and $-L\leqslant M_L\leqslant L$, and we find and examine states that survive the collapse. This opens up the possibility of realizing a metastable system with overcritical numbers of bosons in a ground state with angular momentum $L\neq0$. The multi-orbital mean-field theory predictions about the existence of fragmented metastable states with overcritical numbers of bosons are verified and elucidated at the many-body level.
The descriptions of the total angular momentum within the mean-field and the many-body approaches are compared.

\end{abstract}

\pacs{03.75.Hh, 03.65.-w, 03.75.Kk, 05.30.Jp}

\maketitle

\section{Introduction \label{Introduction}}

Attractive trapped Bose-Einstein Condensates (BEC), since their first realization \cite{Bradley1995, Bradley1997}, have gained increasing attention, due to the interesting phenomena they exhibit \cite{Cornish2000, Sackett1999}. In three dimensions a salient feature is the collapse of the condensate when the (negative) interacting energy per particle, is too large to be compensated by the kinetic energy. The attraction brings the bosons so tightly close that the spatial extension of the wave function of the system shrinks to a point and the condensate eventually implodes \cite{Experiment_Collapse, Controlled_Collapse, Ruprecht_collapse, Collapse-supernova, Collapse-BEC}. However, in trapped gases, \textit{metastable} states, i.e., states that remain stable for some finite time, exist.
The metastability of the condensate is a critical phenomenon: if at a given interaction strength $\lo$, the total number of the bosons exceeds a critical value $\nc$  or vice versa (see for example \cite{Fetter}), the condensate will collapse.

Much work has been devoted to exploring the ground metastable state of attractive BECs and its properties (see, for instance, \cite{Bertsch1999, Jackson2000, Stoof1997, Barberan2006, Dagnino2009}). Recent experiments have revealed new phenomena in attractive BECs that seem to go beyond the ground metastable state. In particular, it was found that states with over-critical number of bosons exist \cite{Cornish2006}. It is natural to assume that excited states of the attractive BECs are involved. Furthermore, disagreements have been reported (see, e.g., \cite{Controlled_Collapse}) between the experiments and the predictions of the Gross-Pitaevskii (GP) theory on the critical value of the attraction strength, where the gas collapses. This motivates us to theoretically study excited states of attractive BECs. We go beyond the standard Mean-Field (MF) theory in the expectation of a thorougher understanding of the role of excited states in the stability of attractive BECs.

To describe the statics and dynamics of a condensate one can adopt a mean-field (MF) approach, such as the famous Gross-Pitaevskii (GP) theory. The starting point of the GP description of a condensate is that all the bosons of the system reside in one and the same one-particle function (orbital) $\phi_0(\r)$ and hence the wave function $\Phi_0$ of the system is merely a product of this prototypal orbital:
\begin{equation}
\label{Phi_GP}
\Phi_0(\r_1,\r_2,\ldots \r_N)=\phi_0(\r_1) \phi_0(\r_2)\ldots \phi_0(\r_N),
\end{equation}
where $N$ is the number of particles.
The expectation value of the system's many-body (MB) Hamiltonian evaluated with this trial function is $E[\Phi_0]=\langle\Phi_0|\hat H|\Phi_0\rangle$. The orbital that minizes this expectation value is the optimal orbital and is found to satisfy a non-linear, partial differential equation, the famous Gross-Pitaevskii equation \cite{Gross, Pitaevskii}.

However, a GP ansatz is not, by definition, capable of describing fragmentation phenomena. The relaxation of the assumption that all bosons are condensed in $\phi_0(\r)$ has been found to lead to fruitful results: energetically favourable fragmented states \cite{BestMeanField, 3Fragm, Properties_of_fragmented_repulsive_cond}, excited metastable states with overcritical number of bosons \cite{cederbaum:040402}, fermionized states and new Mott-insulator phases \cite{fermionization, zoo}. Under this generalized MF approach the wave function of the system is rewritten as:
\begin{equation}
\label{Phi_BMF}
 \Phi(\r_1,\r_2,\ldots,\r_N) = \hat{\mathcal S} \phi_1(\r_1) \phi_1(\r_2)\ldots \phi_1(\r_{n_1}) \phi_2(\r_{n_1+1})\ldots \phi_2(\r_{n_1+n_2}) \ldots \phi_M(\r_N) ,
\end{equation}
where $\hat{\mathcal S}$ is the symmetrizing operator. In such a description and in contrast to the GP case, $M\geq1$ orbitals are allowed to be occupied by bosons. Within this multi-orbital Best Mean Field (BMF) approach \cite{BestMeanField} the orbitals $\phi_j$ of Eq. (\ref{Phi_BMF}) as well as their occupations $n_j$ are calculated \textit{self-consistently}, so as to minimize the total energy of the system. More specifically, the orbitals $\phi_j$ are found to satisfy a system of $M$ coupled non-linear differential equations, in place now of the GP equation. The wave function of the system [Eq. (\ref{Phi_BMF})], i.e., the symmetrized product of $M$ different orbitals is sometimes called `permanent', since it can be regarded as the permanent of a matrix (in direct analogy to the Slater determinant). A condensate, described by Eq. (\ref{Phi_BMF}) is $-$ generally $-$ a \textit{fragmented} \cite{Nozieres1982,Nozieres} condensate, since the reduced one-particle density matrix will give `signatures' in more than one eigenvalues. So, within the BMF framework fragmented states are well described and a GP state arises as an extreme case, where none but one orbital is occupied by all $N$ bosons. Some relevant work to the BMF approach, in favour of or against fragmentation of bosonic systems, can be found in Refs. \cite{Mueller, Fragmented_GS_Spin-1,Fragmented_2D,Ground_band,Spin-1/2-fragmented, Fragmented_MB_ground_states, Absence, Fragility_of_Frag, Saito}.

Going one step further, we write the wave function (\textit{ansatz}) of the system as a linear combination $\Psi=\sum_i C_i \Phi_i$ of different states $\Phi_i$ (permanents), which are taken from a set $\{\Phi_i\}$ of orthonormal permanents $\Phi_i$, each one describing a condensate (fragmented or not) of $N$ particles. This set $\{\Phi_i\}$, which spans our configuration space, consists of all the permanents that result by distributing, in all possible ways, the $N$ bosons over the $M$ orbitals $\phi_j$. Such a state $\Psi$ is known as a \textit{configuration interaction (CI)} expansion \cite{ModernChem}. In contrast to the BMF, a CI state $\Psi$ can further describe purely MB phenomena, such as depletion and fluctuations of the states. The coefficients $C_i$ of the above expansion as well as the orbitals $\phi_j$ are determined variationally. Owing to the detailed analysis of Ref. \cite{CIE} there is a formally defined \textit{Multi-Configurational Hartree} method for \textit{Bosons} (MCHB) and an efficient way to determine the ground and excited states and their energies, i.e., the whole spectrum of a given Hamiltonian in a given configuration subspace of $\{ \Phi_i\}$. The expansion coefficients $\{ C_i\}$ are determined from the diagonalization of the respective \textit{secular} matrix \cite{ModernChem} while the one-particle wave functions $\phi_j$ are obtained by solving a system of $M$ coupled non-linear (integro-) differential equations \cite{CIE}. The MCHB theory and its time-dependent counterpart have been successfully applied to a range of problems of one-dimensional ultracold boson gases, predicting various new phenomena \cite{RoleOfExcited, Bos_Jos_Junc, fragmenton, fragmenton2, Grond2009, Grond2009b}. However, it is, to this day, not feasible to exactly solve the MCHB equations in three-dimensional problems. To overcome this difficulty in the present work, as will be later explained, we implement a restricted version of the MCHB theory, namely a CI expansion of permanents, built over a one-parametric one-particle functions set.

The complexity of a problem, in the framework of MCHB, depends firstly on the number $M$ of the one-particle functions, that are to be determined variationally and secondly, on the symmetries of the system, which in general reduce the size of the configuration space. Since a \textit{complete} configuration space would be infinite-dimensional, the space of the orbitals (and hence that of the permanents) has to be truncated and limited down to a relatively small number $M$, so that real calculations can be performed. Besides this limitation of the configuration space, we imply another constraint on the one-particle wave functions $\phi_j$; we suppose that the one-particle states can be well approximated by wave functions $-$ \textit{ans\"atze} $-$ that are completely known, upon some real parameters $\s_i$. In other words, we fix \textit{a priori} the solutions of the system of $M$ coupled non-linear differential equations to $M$ mono-parametric families of complex wave functions and we then look for the values of $\s_i$ that `optimize' the solution, i.e., values that assign an extremal value to the energy functional. These ans\"atze \cite{Fetter, cederbaum:040402,Variational_ground_state} are taken in our case after the (exact) solution of the corresponding non-interacting system, i.e., they are scaled Gaussians and inherit thus the symmetries of the original system.

In an earlier relevant work \cite{Absence} Elgar\o{}y and Pethick have derived and used a two-mode MB Hamiltonian, borrowed from the nuclear physics Lipkin model, to determine the ground state of an attractive trapped Bose gas. The modes correspond to the s-orbital $Y_{00}$  and the p-orbital $Y_{10}$ and the Hamiltonian matrix is constructed over the set of permanents $|\vec n\rangle=|n_0,N-n_0\rangle$, where $n_0$ bosons reside in the s-orbital and $N-n_0$ in the p-orbital, $N$ being the total number of particles. Then, by rewriting the Hamiltonian in terms of a \textit{quasispin operators} $\hat J_z,\hat J_+,\hat J_-$ they calculated the population at each mode, in the ground state. All the configurations $|\vec n\rangle$ are eigenstates of the quasispin operators with quantum numbers $J=N/2$ and $J_z=\frac{1}{2}(N-2n_0)$. The ground state, in the range of the parameters where it is not collapsed, was found to be not fragmented. However the authors did not examine excited states, which as we shall show in our work, can carry angular momentum (or `non-minimum quasispin' in the case of \cite{Absence}) and are metastable fragmented states that survive the collapse. Still, the work of Ref. \cite{Absence} has stimulated the present extended and more complete study. By including \emph{all} the p-orbitals in our configuration space, we are able to write the wave function of the system as eigenfunction of the true total angular momentum operators and hence restore the symmetries of the problem.

A set of relevant published works, where ultracold bosonic systems are examined with methods beyond the MF approach, includes Refs. \cite{Barberan2006, Dagnino2009, Bertsch1999,Jackson2000}. However they pertain to (true or quasi-) two-dimensional systems, where the description of the angular momentum basis is fairly different and simpler than the analysis on a fully three-dimensional system that we present here. In addition, they do not examine the stability of the system with respect to the fragmentation of ground or excited states.

To render our MB method more efficient we should take all the symmetries of the problem into account. The one-particle functions set $\mathcal{M}$ that we use, i.e., the set of the $\s_i$-orbitals, see Eqs. (\ref{orbitals})-(\ref{orb1}) below, consists of functions that have definite orbital angular momenta, $m_l$ and $l$, as well as parity (symmetry under spatial inversion). However one is more interested in the symmetries of the MB states $\Phi$, since these will directly reduce the size of the configuration (Fock) space.

We principally aim at investigating how the total angular momentum affects the stability and fragmentation of the system. To achieve this we first have to answer on \textit{what are the MB states with definite total angular momentum}. We, therefore, define the MB operators $\hat L^2, \hat L_z$ and their action on the permanents $\Phi_i$. We then look for a $\{\overline{\Phi}_i\}$-basis of MB states that are common eigenstates of $\hat L^2, \hat L_z$. Once the new basis $\{\overline{\Phi}_i\}$ is known we can rewrite the state $\Psi$ and the Hamiltonian of the system on this basis for given eigenvalues $L, M_L$, i.e., over states with the same symmetry. In such a way the size of the new, rotated basis set $\{\overline{\Phi}_i^L\}$, with the index $L$ meaning hereafter that the members of this set have the same angular momentum quantum numbers, is significantly smaller than the original $\{\Phi_i\}$ and the calculations are further facilitated (see also Appendix \ref{appen1}).
We will show that a general state $\overline{\Phi}_k$, with definite angular momentum $L, M_L$, is a quantum MB state, i.e., a non MF state. 

We should also stress the relevance of the `yrast' lines to the present work. The term `yrast' state (or level) has been coined to describe the lowest-in-energy states, for a given angular momentum, first in the context of nuclear physics \cite{Grover1967} and much later in the physics of ultracold Bose gases \cite{Mottelson1999}. Herein we do explore the yrast states of attractive systems but we, also, look at the excited, i.e., non yrast states, for given $L, M_L$. However, the presence of attraction induces a subtle feature. Which state (ground, first excited, etc.) is accretided with the term `yrast' will in principal change, as the absolute value of the interaction strength increases and the lowest-in-energy states start to collapse.

 An equally important goal is to examine the stability and the properties of the ground and excited states of different angular momenta of systems of various $\lo$ and $N$. To do so, we employ the natural orbital analysis; the findings strongly support that states with angular momentum different than zero [large or not, depending on the quantity $\lo (N-1)$]  can exist, with a total number of particles well above the critical number of particles $\ncgp$, as calculated from the GP theory. We verify, therefrom, the predictions of the BMF of Ref. \cite{cederbaum:040402}, that fragmented excited states exist and survive the collapse and we explain these features at the MB level.

The structure of the paper is the following. In Sec. \ref{Prelim} we introduce our theoretical approach to (stationary) quantum bosonic gases and we define the MB and one-particle states that form the configuration spaces. In Sec. \ref{Angular} we give the expression for the total angular momentum operator, we derive a MB angular momentum basis, and we show how this partitions the configuration space. In Sec. \ref{natural_orbital_analysis} we define the main tools of the natural orbital analysis of the MB states. In Sec. \ref{Applications} the main results of this work, for systems of $N=12,60$ and $N=120$ are presented;  in Sec. \ref{appl_1} MB states belonging to the same subspace $L=0$ are compared, while in Sec. \ref{appl_2} we examine ground states of different $L$-subspaces. In Sec. \ref{appl_3} we further investigate the properties, namely fragmentation and variance of the expansion coefficients, see Eqs. (\ref{Psi}) and (\ref{taunorm}) below, of the previously found metastable MB states. In Sec. \ref{comparison} we study the overall impact of the angular momentum on the state of the system with respect to its collapse, and we compare the role of the angular momentum within the MB and the MF theories. Last, Sec. \ref{Outlook} summarizes our results and provides concluding remarks. A set of relevant derivations are given in Appendices \ref{appen1} and \ref{appen2}.

\section{Theory  \label{Theory}}

\subsection{Preliminaries and basic definitions}
\label{Prelim}

We consider a system of $N$ identical spinless bosons of mass \textit{m} confined by an external time-independent potential $V(\r_i)$, and interacting with a general two-body interaction potential $W(\r_i-\r_j)$, where $\r_i$ are the space coordinates of the $i$-th boson.

The Hamiltonian of the system is:
\begin{equation}
 \hat{H}=\hat h+\hat W,
\end{equation}
with $\hat h =\sum_i^N\hat h(\r_i)=-\frac{\hbar^2}{2m} \sum_i^N \hat \nabla^2_{\r_i}+\sum_i^N\hat V(\r_i)$ and $\hat W=\sum_{i<j}^NW(\r_i-\r_j)$ the many-body interaction operator. In the present work we choose $\hat V(\r_i)=\hat V(r_i)$, i.e., the trap potential has spherical symmetry. For the interaction operator we will use the common delta function $\delta(\r-\r')$ representation, $\hat W(\r-\r')=\lo \delta(\r-\r')$, where the parameter $\lo$ measures the strength of the interparticle interaction. This parameter is proportional to the s-wave scattering length $\alpha_s$ and takes on negative values for attractive interaction. Precisely, $\lo=4\pi \alpha_s \sqrt{\frac{\hbar}{m\om}}$, where $\om$ is the frequency of the trapping potential.
The time-independent MB Schr\"odinger equation reads:
\begin{equation}
\label{SE}
 \hat H \Psi = E \Psi,
\end{equation}
where $\Psi=\Psi(\r_1,\r_2,\ldots\r_N)$ is the MB wave function of the system of $N$ interacting bosons and $E$ the eigenvalue of the operator $\hat H$, corresponding to the state $\Psi$.
Even in the simple isotropic case analytic solutions for the MB Schr\"{o}dinger equation are not known. Hence we will approximate the solution $\Psi$ with a MB ansatz as an expansion over known states (permanents) $\{\Phi_i\}$
\begin{equation}
\label{Psi}
 |\Psi\rangle=\sum_i^{N_p} C_i |\Phi_i\rangle,
\end{equation}
where $N_p$ is the total number of permanents used in the expansion and $C_i,~i=1\dots N_p$ the corresponding coefficients. The question that arises is what the set of MB basis functions $\{\Phi_i\}$ consists of. Generally, this includes, as mentioned, all the permanents that result from distributing $N$ bosons over $M$ orbitals.
Readily, these states are those of Eq. (\ref{Phi_BMF}).
In occupation number representation (Fock space representation) the same states take on the form:
\begin{equation}
\label{permanents}
 \phir=|\vec{n}\rangle=|n_1,n_2,\ldots,n_M\rangle.
\end{equation}
Here $n_j$ denotes the respective occupation number of the one-particle functions (orbitals) $\phi_j$.

Mathematically seen, the permanents $\phir$ are the vectors that span the Fock space $\mathcal F$ of all $N$-body wave functions. However it is possible to reduce the size of the `working' Fock spaces, by partitioning the initial space into $\Pi$- and $L$-subspaces, i.e., spaces of permanents with definite parity and angular momentum.
The purpose to do so is twofold; firstly the resulting solution $\psir$ will possess the rotational symmetries of the system and secondly the working $\{\Phi_i^L\}$ spaces are each time much smaller in size than the initial $\{\Phi_i\}$ one. Note that, while this partioning of the configuration space with respect to angular momentum $L$ is trivial in a two-dimensional system, it is not straightforward in the full 3D case, that we examine here.

The MB Hamiltonian $\hat H$ in the $\{\Phi_i\}$-basis is represented as a \emph{secular} matrix $\mathcal H$, with elements: 
\begin{equation}
\label{secularm}
 \mathcal{H}_{i,j} = \langle \Phi_i | \hat H | \Phi_j \rangle.
\end{equation}
By diagonalizing, i.e., solving the equation
\begin{equation}
 \mathcal{H} \mathbf{C}=\mathcal{E}\mathbf{C},
\end{equation}
where $\mathbf C$ is the column vector of the expansion coefficient  $\mathbf C=\{C_1,C_2\ldots,C_{N_p}\}^T$ and $\mathcal E$ the respective eigenvalue, we obtain the energies (eigenvalues) and coefficients (eigenvectors) of the solutions of the system.
Note that the Hamiltonian in Eq. (\ref{secularm}) is evaluated still in the full basis $\{\Phi_i\}$ of permanents.

To complete the picture of our variational solutions, we give the one-particle function basis set, over which the permanents of Eq. (\ref{Phi_BMF}) are constructed. This set of \textit{ans\"atze} consists of the known orbitals that solve the isotropic 3D quantum harmonic oscillator, scaled under a scaling parameter $\s_i$. Precisely, it consists of four orbitals: the ground $l=0$ and the three $l=1$ excited ones, which we scale with two parameters ($\so,\se$), as have been already done in Ref. \cite{cederbaum:040402}. The parameters $\sigma_i$ will determine the shape (width) of the orbitals; their optimal values are such that, for a given set of coefficients $\mathbf C$, the total energy of the system takes an extremum. In this approximative way we restrict the solution of the system to functions that lie inside the monoparametric families of equations, which solve a scaled ordinary Schr\"odinger equation; the solution of a coupled system of nonlinear differential equations (MCHB equations) boils down to the determination of a set of parameters which minimizes the total energy.

In this work much of the analysis of the quantities of the system (energy per particle, occupation numbers, variances of expansion coefficients) is done with respect to the scaling parameters. So for example, we expect to see a (local) minimum in the plot of $E(\so,\se)$ against $\so,\se$ if the system is metastable, while the absence of minimum will signal a collapsing condensate. Moreover, the analysis of the resulting MB states (depletion, angular momentum) is performed at the optimal values of the parameters $\si$, $i=0,1$.

The orbitals that we use, in Cartesian coordinates, have the form:
\begin{equation}
\begin{gathered}
\label{orbitals}
\phi_1(\r)=\varphi_0(x,\so) \varphi_0(y,\so) \varphi_0(z,\so),   \\
\phi_2(\r)=\frac{1}{\sqrt{2}}(\varphi_1(x,\se) \varphi_0(y,\se) \varphi_0(z,\se)+i \varphi_0(x,\se) \varphi_1(y,\se) \varphi_0(z,\se)), \\
\phi_3(\r)=\varphi_0(x,\se) \varphi_0(y,\se) \varphi_1(z,\se), \\
\phi_4(\r)=\frac{1}{\sqrt{2}}(\varphi_1(x,\se) \varphi_0(y,\se) \varphi_0(z,\se) - i \varphi_0(x,\se) \varphi_1(y,\se) \varphi_0(z,\se)), \\
\end{gathered}
\end{equation}
where
\begin{equation}
\label{orb0}
 \varphi_0(x,\s)=\left(\frac{m\om}{\pi \s^2 \hbar}\right)^{1/4} e^{-\frac{1}{2} \frac{m\om}{\s^2 \hbar} x^2}
\end{equation}
and
\begin{equation}
\label{orb1}
\varphi_1(x,\s)=\left[\frac{4}{\pi}\left(\frac{m\om}{\s^2 \hbar}\right)^3\right]^{1/4} x e^{-\frac{1}{2} \frac{m\om}{\s^2 \hbar} x^2}
\end{equation}
are orthonormal orbitals, i.e., $\langle \varphi_i|\varphi_j\rangle=\delta_{ij}$, $i,j=0,1$. Here $m$ is the particle mass and $\om$ is the trap frequency. 
Throughout this work the quantities used are dimensionless, i.e., $\hbar=m=\om=1$.

In terms of spherical harmonics $Y_{l,m_l}$, i.e., under a change of coordinates, the orbitals of Eq. (\ref{orbitals}) are: 
\begin{equation}
\label{orbitalsY}
\phi_{k}(\r)=\varphi_l(r,\sigma_l) Y_{lm_l}(\theta,\phi),
\end{equation}
where $l=0,1$, $-l\leqslant m_l\leqslant l$ and $k\equiv k(l,m_l) = 1+l(l+1)-m_l$.

\subsection{Angular momentum basis \label{Angular}}

It is easy to see that the orbitals of Eq. (\ref{orbitalsY}) constitute a set of common eigenstates of the orbital angular momentum operators $\hat L^2, \hat L_z$ together with the parity (inversion) operator $\hat \Pi: \hat\Pi\Psi(\r)=\Psi(-\r)$, with eigenvalues $l=\{0,1,1,1\}$, $m_l=\{0,1,0,-1\}$  and $\pi=\{1,-1,-1,-1\}$, respectively.

We now want to express the total angular momentum operators at the MB level. For this purpose we switch to second quantization language and introduce the bosonic creation (annihilation) operators $b_i^\dagger$ ($b_i$), associated with the orbital set $\{\phi_i(\r)\}$ and which obey the usual bosonic commutation relations: $b_i b_j^\dagger-b_j^\dagger b_i=\delta_{ij}$.

The total angular momentum operators are (see, e.g., \cite{Schirmer}):
\begin{equation}
\label{L2}
\hat L^2=\hat L_z^2+\frac{1}{2} (\hat L_+ \hat L_- +\hat L_- \hat L_+),
\end{equation}
\begin{equation}
\label{Lz}
 \hat L_z = \sum_{l,m_l} m_l b^\dagger_{l m_l} b_{l m_l},
\end{equation}
\begin{equation}
 \hat L_\pm = \sum_{l,m_l} A(l, \mp m_l) b^\dagger_{l m_l\pm1} b_{l m_l},
\end{equation}
where 
$A(l,m_l)=[(l+m_l)(l-m_l+1)]^{1/2}$ and $b_{lm_l}^\dagger$ ($b_{lm_l}$) creates (annihilates) a boson in the state $\phi_{lm_l}$, with orbital angular momentum quantum numbers $l,m_l$.

Applying Eq. (\ref{L2}) to our basis of Eq. (\ref{permanents}) with $M=4$ we get:
\begin{equation}
\label{L}
 \begin{gathered}
\hat L^2 |\vec{n}\rangle=\left[n_2(n_3+1)+n_3(n_4+1)+n_3 (n_2+1)+n_4(n_3+1)+ (n_2-n_4)^2\right] |n_1,n_2,n_3,n_4\rangle+ \\ 
+2 \sqrt{n_3 (n_3-1) (n_2+1)(n_4+1)} |n_1,n_2+1,n_3-2,n_4+1\rangle+ \\
+2\sqrt{n_2 n_4(n_3+1)(n_3+2)}|n_1,n_2-1,n_3+2,n_4-1\rangle.
\end{gathered}
\end{equation}
It can be easily seen that each permanent [Eq. (\ref{permanents})] is an eigenstate of $\hat L_z$ with eigenvalue $M_L=n_2-n_4$,
\begin{equation}
\label{Lzeiv}
\hat  L_z |\vec n\rangle = (n_2-n_4) |\vec n\rangle.
\end{equation}

But what happens to the eigenstates and eigenvalues of the $\hat L^2$ operator? To answer this, one has to solve the eigenvalue equation:
\begin{equation}
 \mathcal L \mathcal C = \Lambda \mathcal C,
\end{equation}
where $\mathcal L$ is the matrix representation of the operator $\hat L^2$ in the basis of permanents [Eq. (\ref{permanents})] with matrix elements
\begin{equation}
\label{Lmat}
\mathcal L_{i,j}= \langle \vec n_i |L^2|\vec n_j\rangle  ,
\end{equation}
$\mathcal C$ is the column vector of the coefficients $C_i$ and $\Lambda$ the eigenvalue in question.

A unitary transformation $\mathcal U$ will in general rotate the $\Phi$-basis to a new $\overline{\Phi}$ one. In this basis the secular matrix of Eq. (\ref{secularm}) becomes:
\begin{equation}
 \begin{gathered}
\label{secularmr}
 \mathcal{\overline H}_{i,j} = \langle \overline\Phi_i | \hat H | \overline\Phi_j \rangle = 
\sum_{k,l} \mathcal U_{i,k}^\dagger \mathcal{H}_{k,l} \mathcal U_{l,j},
\end{gathered}
\end{equation}
where $\mathcal U_{i,j}$ are the matrix elements of $\mathcal U$. If $\mathcal U$ is simply the matrix of the eigenvectors of $\mathcal L$ then $\overline{\mathcal H}$ takes on the desired total-angular-momentum block-diagonal form.
The above mentioned vector spaces, that the bases $\{\Phi_i\}, \{\overline \Phi_i\}$ span, are homomorphic and can be written as the direct sum of the subspaces $\{\overline\Phi_i^L\}$:
\begin{equation} 
\{\Phi_i\}\cong \{\overline\Phi_i\} =\bigoplus_{L} \{\overline\Phi_i^L\}.
\end{equation}

We have numerically calculated the angular momentum states $\overline{\Phi}_i$ and the matrix transformation $\mathcal U$  that transforms to the new basis $\overline{\Phi}=\mathcal{U} \Phi$ of eigenstates of $\hat L$ and $\hat M_L$, for the cases of $N=12, 20, 60, 120$ bosons. An analytic approach to the same problem of determining the states $\overline{\Phi}_i$ is presented in Appendix \ref{appen2a}.
For selected values of $L$ and $M_L$ we construct and diagonalize the block $\mathcal H^L$ of the secular Hamiltonian matrix $\mathcal{\overline H}$ and find its eigenfunctions $|\Psi^L\rangle$. We use hereafter the index $L$ to stress the fact that the state $\psil$ is an eigenstate of $\hat L^2$ with quantum number $L$. The size of the block $\mathcal H^L$ is found to be $N_p=\frac{N-L+2}{2}$ (see also Appendix \ref{appen1}), with the same number of eigenstates. We index the states $\psili$ with $i$, to denote the ground ($i=1$) and the excited ($1<i\leq \frac{N-L+2}{2}$) states belonging to this block of angular momentum $L$.  When it is not transparent from the context, we will also use $\lambda_{0,cr}^L$ or $\lambda_{0,cr}^i$, to denote the critical value of the interaction strength where the state $\psili$ collapses.

\subsection{Natural orbital analysis \label{natural_orbital_analysis}}

The first order reduced density matrix (RDM) for the state $\psir=\sum_{\vec n} C_{\vec n}|\vec n\rangle$, Eq. (\ref{Psi}), is defined as:
\begin{equation}
\label{1BRDM}
 \rho(\r|\r')= N \int \Psi^\ast(\r',\r_2\dots \r_N) \Psi(\r,\r_2\dots \r_N) d\r_2 d\r_3\dots d\r_N = \sum_{i,j}^M \rho_{ij} \phi_i^{\ast}(\r') \phi_j(\r)
\end{equation}
and the second order RDM:
 \begin{align}
\label{2BRDM}
 \rho(\r_1,\r_2|\r_1',\r_2')=N (N-1) \int \Psi^\ast(\r_1',\r_2',\r_3\dots \r_N) \Psi(\r_1,\r_2,\r_3\dots \r_N) d\r_3\dots d\r_N = \nonumber \\
\sum_{i,j,k,l}^M \rho_{ijkl} \phi_i^\ast(\r_1') \phi_j^\ast(\r_2') \phi_k(\r_1) \phi_l(\r_2),
\end{align}
with $\rho_{ij}=\langle \Psi |b^\dagger_i b_j |\Psi\rangle, \rho_{ijkl}=\langle \Psi |b^\dagger_i b^\dagger_j b_k b_l |\Psi\rangle$ being the elements of these matrices.
The \emph{natural orbitals} $\phi_i^{NO}$ are defined as the eigenfunctions of $\rho(\r|\r')$, i.e., the one-particle functions that diagonalize the right hand side of Eq. (\ref{1BRDM}) and their eigenvalues are known as \emph{natural occupations} $\rho_i$. In our system, the spherical symmetry of the Hamiltonian induces a zero $\mathcal H$-matrix element between states of different symmetry (angular momentum and parity). Hence $\rho_{ij}$ is diagonal and the ansatz orbital-set of Eq. (\ref{orbitals}) coincides with the set of the natural orbitals $\{\phi_i^{NO}\},~i=1,\dots,M$. 

The diagonal elements
\begin{equation}
\label{occupation}
\rho_{ii}\equiv\rho_i=\sum_{\vec n} C_{\vec n}^\ast C_{\vec n} n_i = \langle \hat n_i \rangle,
\end{equation}
$i=1,\dots ,M$, are the expectation values of the number operators associated with the orbitals $\phi_i$ and are, as explained, the natural occupations $\rho_i$ of the respective natural orbitals. We define the \emph{depletion} $d_i$ of the $i$-orbital as:
\begin{equation}
\label{depletion}
d_i=1-\rho_i/N .
\end{equation}
The depletion $d_i$ is an informative quantity which measures the relative number of particles that are depleted from the $i$-orbital. Throughout this work we extensively use the quantity $d_1$ and also refer to it as the s-depletion.
The diagonal elements
\begin{equation}
\label{variances}
\rho_{iiii} =\sum_{\vec n} C_{\vec n}^\ast C_{\vec n} (n_i^2-n_i) = \langle \hat n^2_i \rangle-\langle \hat n_i \rangle
\end{equation}
are related to the \emph{variance} of the distribution of the coefficients $C_i$ of a given state $|\Psi\rangle$, associated with the orbitals $\phi_i$, as:
\begin{equation}
\label{tau}
 \tau_i^2 \stackrel{\text{def}}{=} \langle \hat n_i^2\rangle - \langle \hat n_i \rangle ^2 = \rho_{iiii}+\rho_{ii}(1-\rho_{ii}).
\end{equation}
The norm
\begin{equation}
\label{taunorm}
 |\tau| \stackrel{\text{def}}{=} \sqrt{\ta_1^2+\ta_2^2+\ta_3^2+\ta^2_4}
\end{equation}
gives a measure of the variances \footnote{To be precise, the quantity $\tv$ is the norm of the \emph{standard deviations} $\tau_i$, commonly defined as the square root $\sqrt{\tau_i^2}$ of the variance $\tau^2_i$. To avoid confusion we make clear that we use in this work the term \emph{variance} to refer to the norm $\tv$.} of all four orbitals of a state $|\Psi\rangle$. The $\ta_i$'s and $\tv$ are highly useful quantities, as they measure the fluctuations around the occupation numbers of the natural orbital $\phi_i^{NO}$. In the case of a mean-field state, where there are no fluctuations, $\tv$ is simply zero \footnote{We have numerical indication that in any state $|\Psi\rangle$  the relation $\ta_2=\ta_4$ holds. For the special case of zero angular momentum (spherically symmetric $|\Psi\rangle$) we get further $\ta_{2,4} =\sqrt{2 \ta_3^2+\langle \hat n_3\rangle (\langle \hat n_3\rangle+1)}/\sqrt{3}$. So, even in the spherically symmetric states with $\langle \hat n_2\rangle=\langle \hat n_3\rangle=\langle \hat n_4\rangle$ the variances are not equal, as one might expect, but instead related through the above equation.}.

\section{Many-Body Results \label{Applications}}

In this section we implement the many-body method described above, for systems of trapped ultra-cold gases. We present and discuss calculations regarding systems of $N=12, 60$ and $120$ bosons, embedded in a spherically symmetric trap. First, the interaction strength $\lo$ is chosen each time such that the product $|\lo| N$ is kept fixed to the value $10.104$ \footnote{The reader should not get the impression that this choice conflicts the $|\lo|(N-1)$ scaling in the subsequent analysis of Secs. \ref{Applications} and \ref{comparison}. The value $|\lo|N=10.104$ is chosen just to ensure that the GP ground state of systems of \emph{any} number of bosons $N$ will be collapsed. We would not have obtained fairly different results if the choice $|\lo|(N-1)=const.$ was made instead.}. This choice will permit a direct comparison of our results to those of Ref. \cite{cederbaum:040402}, where an attractive system of $|\lo| N=10.104$ was also examined. Later on, also other values of $|\lo| N$ are considered for the shake of completeness. In the following we examine states of definite angular momentum $L, M_L$ and positive parity $\Pi$ only. The latter makes the total angular momentum of each state increase at an even step, i.e., $L=0,2,4,\dots, N$.

\subsection{Ground and excited states of the `block' \textit L=0 \label{appl_1}}
\label{groundst}

\begin{figure}[ht]
	\centering
	\includegraphics[scale=0.6]{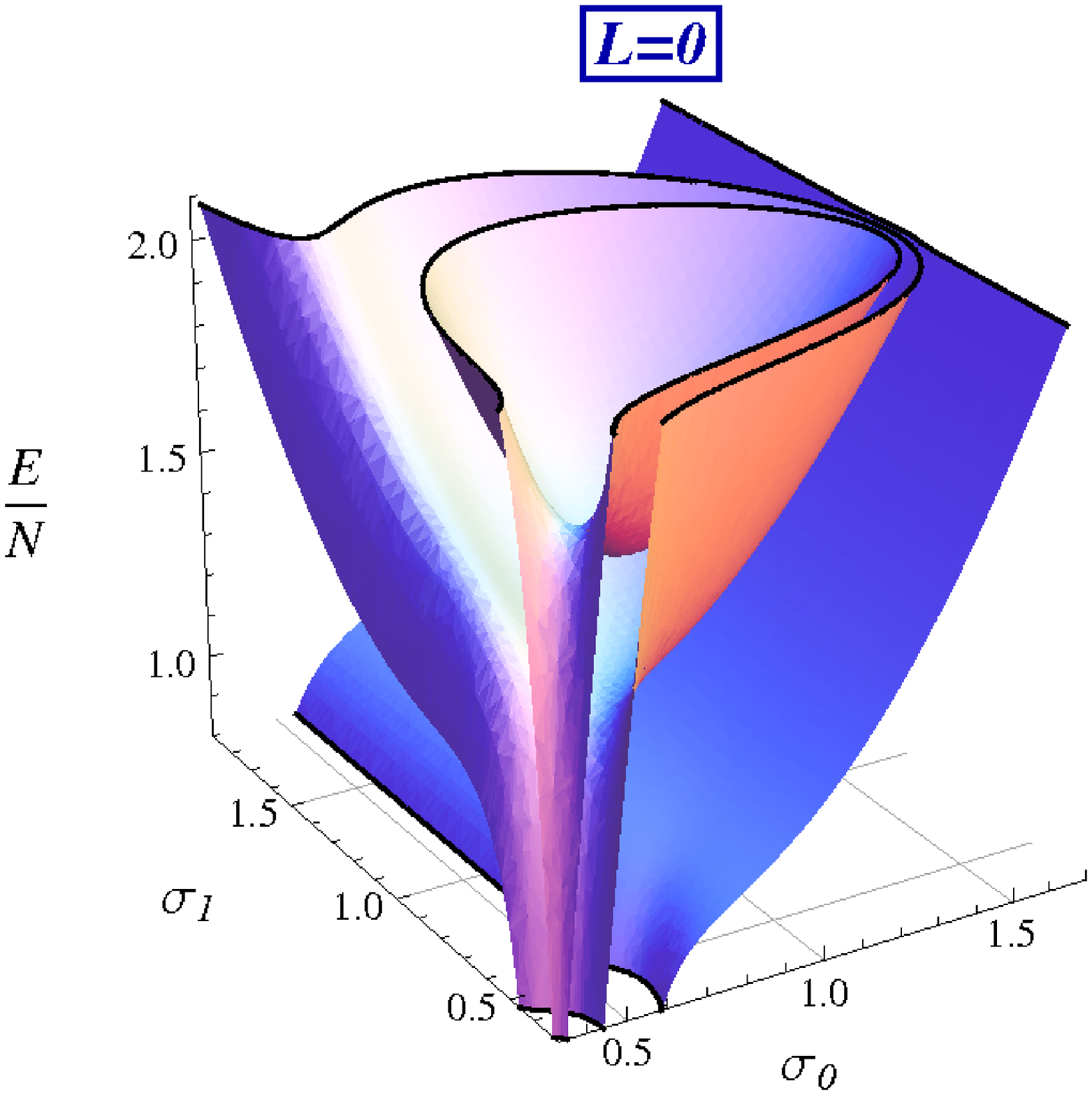}
		\caption{(Color online) Energy landscape $\e(\so,\se)$ for a system of 
		$N=120$ bosons, $\lo=-0.0842$. Shown are energy
		surfaces of the ground ($i=1$), the $i=20$ and the $i=30$ excited MB states, all with 
		$L=M_L=0$. The lowest surface corresponds to the ground state of the coherent system,
		where almost all 120 bosons reside in the s-orbital. It exhibits no minimum and hence the system
		collapses. The middle surface barely exhibits a minimum, at $\so=0.72,\se=0.70$,
		with barrier height $h=3.83\cdot10^{-3}$. The natural occupations are 
		$\rho_1=80.82$, $\rho_2=13.06$, $\rho_3=13.06$ and $\rho_4=13.06$.
		In the third surface a clear minimum in the energy, $\e_0=1.60$, is shown, at $\so=0.82, \se=0.81$, 
		with barrier height $h=0.23$.
		The occupation numbers, at this point, are $\rho_1=62.13, \rho_2=19.29, \rho_3=19.29, \rho_4=19.29$.
		All quantities are dimensionless.
		 \label{energies_L0}}
\end{figure}

For each of the above systems we examine states with definite angular momentum. We first calculate the energy per particle $\e =\mathcal E/N$ of those states, as a function of the variational parameters $\sigma_0, \sigma_1$ of the orbitals [see Eqs. (\ref{orb0}) and (\ref{orb1})]. We then look for the minimum $\e_0$ of the energy with respect to these parameters. As mentioned earlier, the total absence of a minimum indicates unbound (total) energy and a collapsing system. Namely, as $\so,\se\rightarrow0$ the orbitals of Eq. (\ref{orbitals}) contract to pointlike distributions. When existent, the minima are expected to be local only; an energy barrier separates the metastability from the collapse regions. The shape of the energy barrier determines the tunneling time of the system through this barrier and $-$ generally $-$ the higher the barrier is, the longer the system is expected to survive in this state.
The variation of $\so,\se$ takes place over states of the same symmetry and hence the surfaces ought not to cross (see Fig. \ref{energies_L0}. See also the theoretical discussion on non-crossing of energy surfaces in  \cite{Energy_crossing,Behaviour_of_Eigenv} and references therein). Notice that, owing to the attractive interparticle interaction, the wave function of the system has to be spatially shrunk, compared to that of the non-interacting system; indeed the optimal scaling parameters of the orbitals, are always found to obey $\so,\se < 1$.

The first system studied is that of $N=120$ bosons, with attractive interaction of strength $\lo$=-0.0842. The energies per particle $\e(\so,\se)$ of three distinct states of this system are collectively presented in Fig. \ref{energies_L0}. We first pick the state with quantum numbers $L$=0 and $M_L$=0 of the operators $\hat L^2, \hat L_z$, respectively.  We find that the ground state is collapsed (lowest surface in Fig. \ref{energies_L0}). As the introduction of Sec. \ref{Introduction} suggests, we expect to find excited, fragmented states that can survive this collapse. Indeed, an examination of the spectrum of the states of the Hamiltonian of Eq. (\ref{secularm}) reveals that the $|\Psi^{L=0}_{i=20}\rangle$ excited state is the first to demonstrate a minimum in the energy (middle surface in Fig. \ref{energies_L0}) and this makes it the yrast state, for this $\lo$ and $L$. The optimal values of the sigmas are $\sigma_0=0.72,\sigma_1=0.70$, the minimum energy per particle for these values of sigmas is $\e_0=1.37$ and the s-depletion is $d_1=0.33$. However the energy barrier, that prevents the system from collapse, is extremelly low, $h\sim10^{-3}$, making the state only marginally metastable. On the other hand, the $|\Psi^{L=0}_{i=30}\rangle$ excited state of the system exhibits a clear minimum (energy barrier height $h=0.23$), with energy per particle $\e_0=1.60$ and s-depletion $d_1=0.48$ at the optimal values of the sigmas $\sigma_0=0.82,\sigma_1=0.81$ (upper surface in Fig. \ref{energies_L0}).

For $L=0$ a metastable fragmented state can decay by two channels. The first, as mentioned above by tunneling through the barrier. The second, by coupling to lower surfaces with the same $L=0$ which do not have a minimum. Since all these surfaces do not have a minimum are energetically far below, the coupling between them is not expected to induce a quick collapse. 
Consequently, metastable excited states with $L=0$, with parameter values tuned at the collapse region of a GP state, exist at higher energies.

\subsection{Ground states for various angular momenta \textit{L}  \label{appl_2}}

\begin{figure}
\centering
\vspace{-120pt}
\includegraphics[scale=0.7]{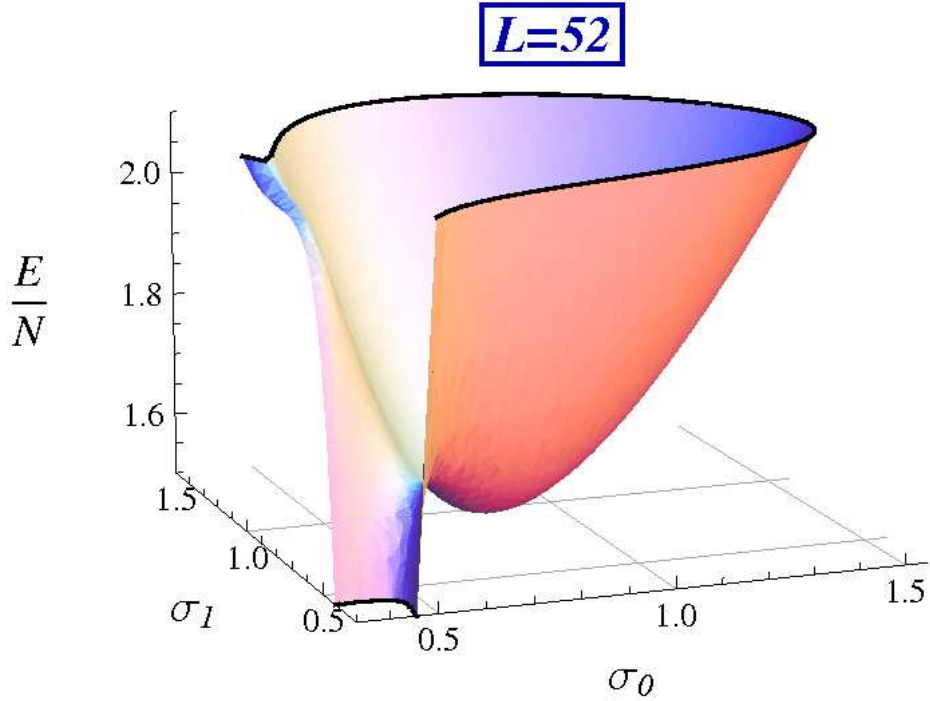}
\vspace{-70pt}
\caption{(Color online) For the same system as in Fig. \ref{energies_L0}, i.e., $N=120, \lo=-0.0842$ we plot the energy surface of the ground state with angular momentum $L=52, M_L=0$. Clearly there is a minimum in the surface, which manifests metastability of the system. Contrarily, when $L=0$ (Fig. \ref{energies_L0}) the ground state is found to be collapsed. The minimum energy per particle is $\e_0=1.59$, at point $\so=0.82, \se=0.81$ and the occupation numbers $\rho_1=62.30, \rho_2=14.30, \rho_3=29.10, \rho_4=14.30$. All quantities are dimensionless.}
\label{energy_l52}
\end{figure}

Next, we perform the same analysis as in section \ref{groundst} for the system of $N=120$ bosons, this time over states of significantly higher angular momentum. Precisely we choose states with $L=52, M_L=0$. We recall that the maximum allowed quantum number for the total angular momentum, within the present analysis, is $L_{max}=N=120$. We want to compare the stability and the properties of the two systems, namely that of $L=0$ to that of $L=52$. The energy surface $\epsilon(\so,\se)$ as a function of the scaling parameters $\so, \se$ is plotted in Fig. \ref{energy_l52}. A clear minimum can be seen, at $\so=0.82,\se=0.81$ and $\e_0=1.59$ manifesting a metastable ground state with $L=52$, for the same system whose $L=0$ ground state is found to be collapsed.
We should stress here that the state we examine is the \emph{lowest} in energy state of this $L$ and so this makes it the ground (yrast) state of the problem.

\begin{figure}
\centering
\vspace{-30pt}
\hspace{-10pt}
\includegraphics[scale=0.38]{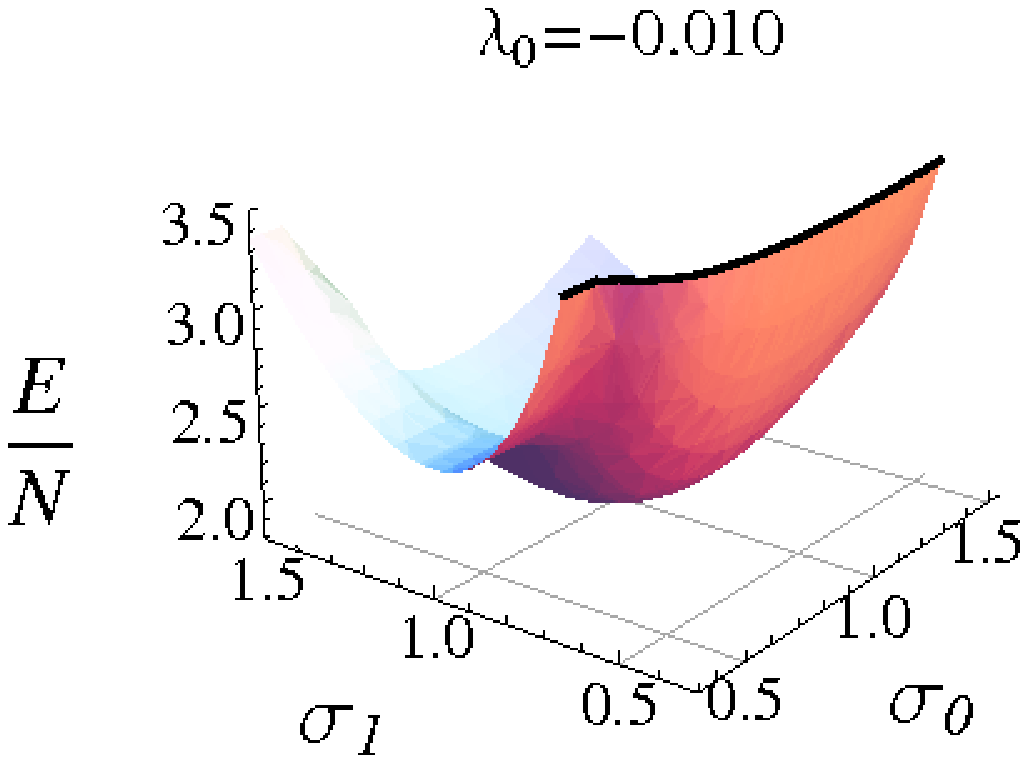}
\hspace{-40pt}
\includegraphics[scale=0.38]{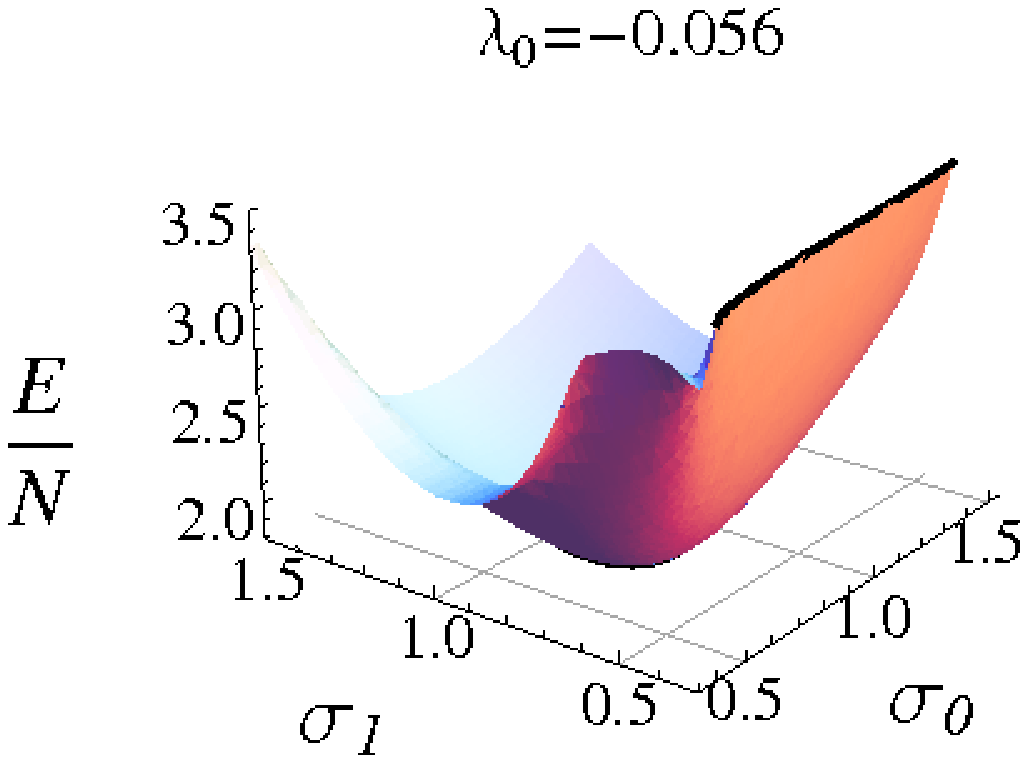}
\hspace{-40pt}
\includegraphics[scale=0.38]{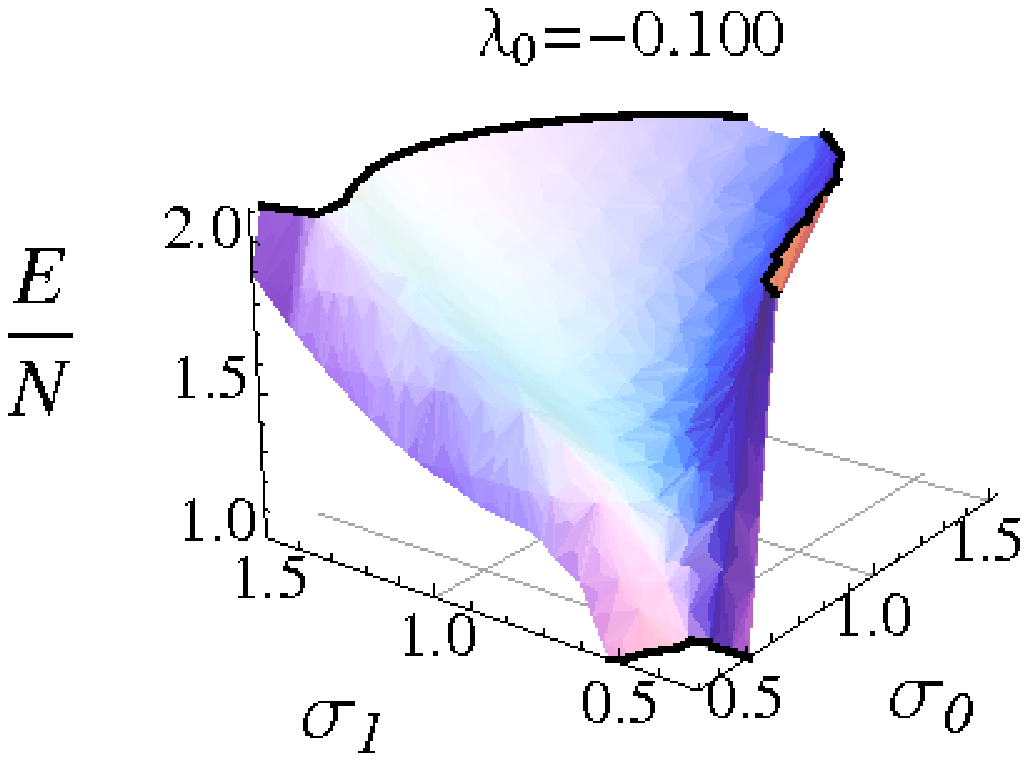}
\vspace{-30pt}
\caption{(Color online) Energy surfaces for the state of Fig. \ref{energy_l52} for values of interaction strength $\lo=-0.010, \lo=-0.056$ and $\lo=-0.100$. The first two surfaces exhibit minima, i.e., metastability, while the third one does not and hence the system collapses. See text for more details. All quantities are dimensionless.}
\label{energy_l52_lambdas}
\end{figure}

In Fig. \ref{energy_l52_lambdas} we plot the energy surface of the same state, examined above, for different values of the interaction strengh, $\lo=(-0.010,-0.056,-0.100)$. For small values of $\lo=(-0.010,-0.056)$ the energy surface exhibits a clear minimum, with its energy barrier being higher than in the case of $\lo=-0.0842$. In the third picture, the energy surfaces shows no minimum, meaning that this state is collapsed, though the critical value $\lambda_{0,cr}^{L=52}$ is much higher than the corresponding $\lambda_{0,cr}^{L=0}$ of the $L=0$ state.

Following Fig. \ref{energies_L0}, a plot of energy surfaces of ground (yrast) states, $i=1$, with different angular momentum and hence different stability behaviour, would be intuitive. If one would plot the energies of the group of ground states $|\Psi^{L=0}_{i=1}\rangle$, $|\Psi^{L=38}_{i=1}\rangle$ and $|\Psi^{L=58}_{i=1}\rangle$ on the ($\so, \se$) plane, they would see that the resulting graph would look very much like that of Fig. \ref{energies_L0}. This means that the energy surfaces of the pairs of states $|\Psi^{L=38}_{i=1}\rangle$ and $|\Psi^{L=0}_{i=20}\rangle$ as well as $|\Psi^{L=58}_{i=1}\rangle$ and $|\Psi^{L=0}_{i=30}\rangle$ are almost the same, for all $\so, \se$. This coincidence is not an accident. Indeed, as we shall show later, one can find states that are very close $-$ almost degenerate $-$ in energy but have different angular momentum quantum number $L$ (see discussion at the end of Sec \ref{fragmentation}).

As a direct generalization of the above, we can say that if, for some $\lo$ the ground state with $L=0$ of the $N$-boson system is collapsed, then there will be a ground state with angular momentum $L>0$, large enough so as to survive the collapse. Further, if the interaction strength is increased, past some new critical value, this state will also collapse.

\subsection{Analysis and structure of the energy surfaces \label{appl_3}}

To thoroughly analyze the properties of the MB states we examine the findings of the previous sections under the light of the natural orbital analysis and the use of RDMs. For given ground and excited metastable states $|\Psi\rangle$ we want to answer on: ($i$) what the natural occupations are, ($ii$) how much fragmented the states are and ($iii$) how much they deviate from MF states, in a range of the parameters $\so, \se$ as well as $\lo, L, M_L$. The systems examined in this section consist of $N=12$ and $N=60$ bosons and the interaction strength is set to $\lo=-0.842$ and to $\lo=-0.1684$, respectively.

\subsubsection{Fragmentation \label{fragmentation}}

As mentioned, due to the symmetry of the Hamiltonian, the natural orbitals of Eq. (\ref{1BRDM}) coincide with those defined in Eq. (\ref{orbitals}), for all $\lo, \so, \se$.  It is interesting to see how the occupations $\rho_i$, defined in Eq. (\ref{occupation}), of the ground and excited metastable states of definite $L$, vary in the ($\so, \se$) plane or change with  $\lo$. Unlike $\rho_{2},\rho_3$ or $\rho_4$, the quantity $\re$ (or $d_1$) is invariant $-$ for given $L$ $-$ under changes of the quantum number $M_L$ of the operator $\hat L_z$. Furthermore, as long as solely ground states are considered, i.e., $i=1$, $\re$ determines the total angular momentum $L$. These properties make $\re$ a quite informative and representative quantity of the state $|\Psi\rangle$.

\begin{figure}
\centering
\includegraphics[scale=.65]{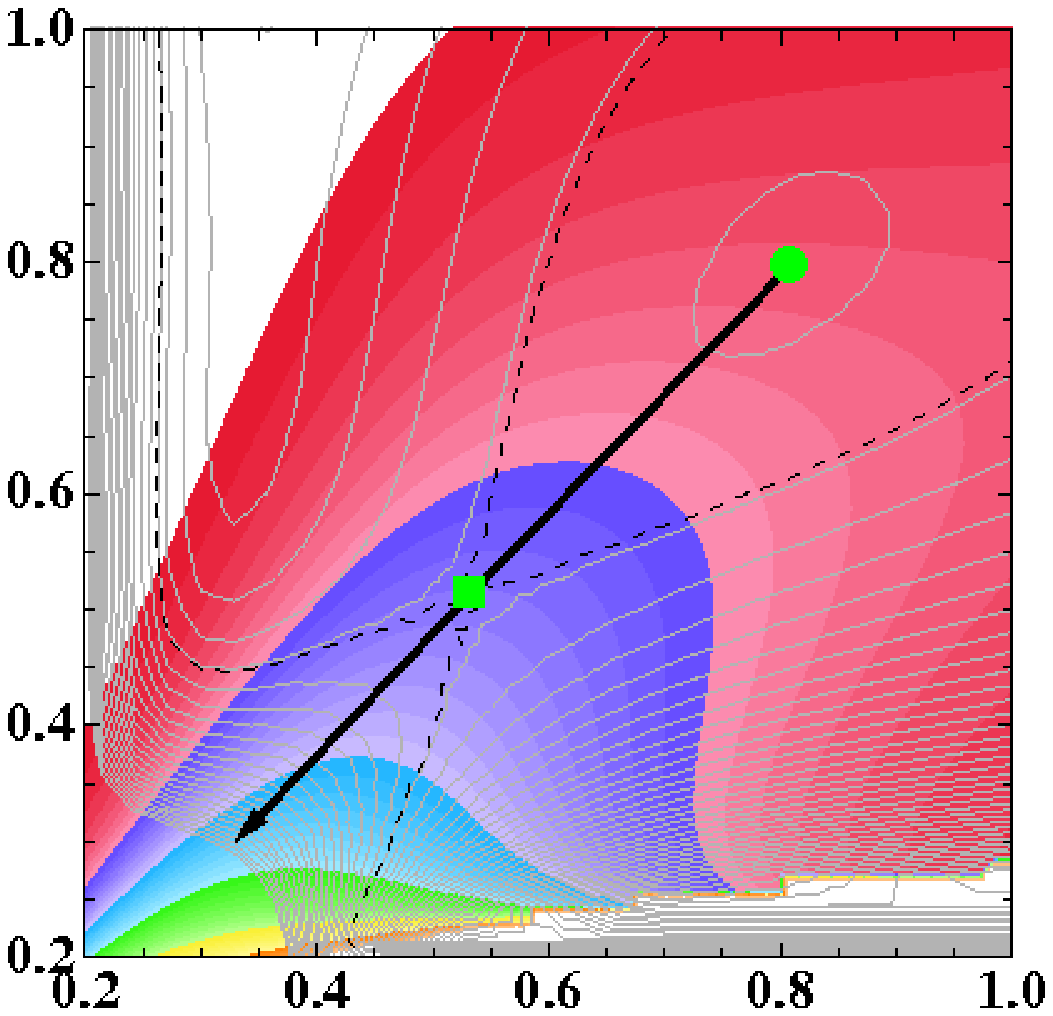}
\hspace{10pt}
\includegraphics[scale=0.123]{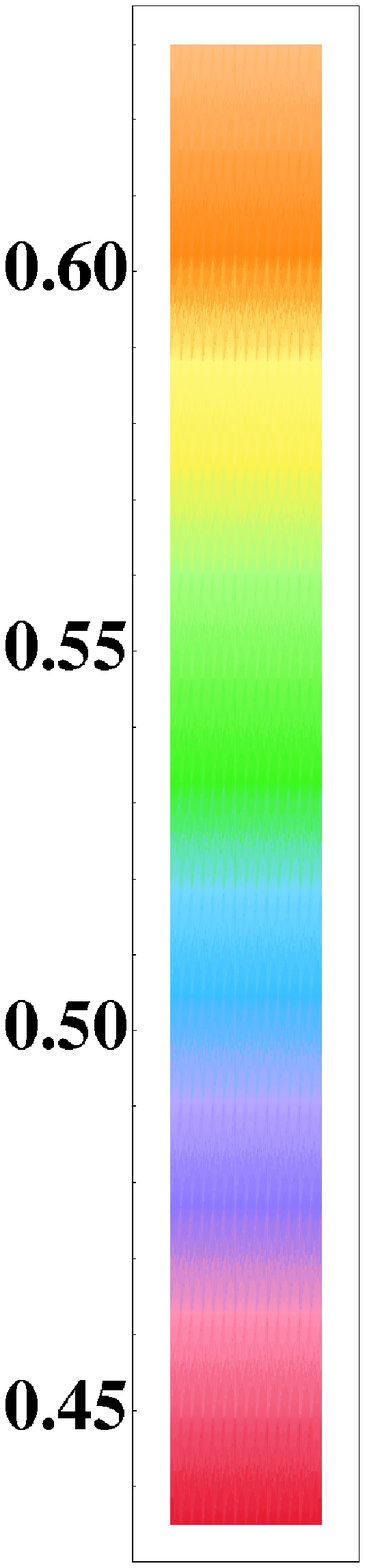}
\caption{(Color online) Change in the s-depletion of the condensate ($d_1=1-\rho_1/N$) on the ($\so,\se$) plane for a metastable ground state with $L=26, M_L=2$ of an $N=60$, $\lo=-0.1684$ system. Plotted are contour lines of fixed $\rho_1= const.$. The minimum of energy (green dot in the plot) is $\e_0=1.55$ at $\so=0.81,\se=0.79$ and the s-depletion is $d_1=0.45$. The saddle point (green square) on the energy surface gives the maximum energy that a metastable state can have. Along the `collapse path' (arrow) $d_1$ increases; by moving the system $-$ over the energy barrier $-$ it becomes more and more fragmented. All quantities are dimensionless.}
\label{Occupations-sig_60}
\end{figure}

For a system of $N=60$ bosons in the ground metastable state with $L=26, M_L=2$, we calculate the depletion of the $\phi_1$-orbital (s-depletion $d_1$), see Eq. (\ref{depletion}), as a function of the parameters $\so, \se$. In Fig. \ref{Occupations-sig_60} we plot the contour lines $\rho_1=const.$, versus the parameters $\so, \se$. The energy landscape of this particular state, for this choice of parameters would look very much like the one of Fig. \ref{energy_l52}. 
To allow a monitoring of the energy surface, we also plot in Fig. \ref{Occupations-sig_60} the contours (light grey) of constant energy $\epsilon$. The dashed line is the highest-in-energy contour that corresponds to a metastable state. It splits the graph into four parts; in the upper right one the `trajectories' are bounded, while they are not in the other parts of the space (hyperbolic trajectories). Thus it resembles a \emph{separatrix} of a phase space, whose trajectories meet asympotically only in a saddle point. 
The energy per particle has a local minimum $\e_0=1.55$ at $\so=0.81, \se=0.79$ and at this point the s-depletion is found to be $0.45$, i.e., $55\%$ of the particles of the system are excited to the orbitals $\phi_j,~j=2,3,4$. 

Of special interest is also the change of the s-depletion as the system moves towards the collapse. To make this evident we have plotted on Fig. \ref{Occupations-sig_60} an arrow marking the `collapse path', i.e., the line that connects the minimum (green dot) with the saddle point (green square) of the energy surface, i.e., the maximum of the energy barrier. Along this path the system moves over the energy barrier towards collapse and it crosses contours of different $\re$; as collapse takes place the s-depletion of the state increases. We note that for large values of the scaling parameters, i.e., $\so,\se\gg1$ the s-depletion remains practically unchanged.

\begin{figure}
\centering
   \includegraphics[scale=.60]{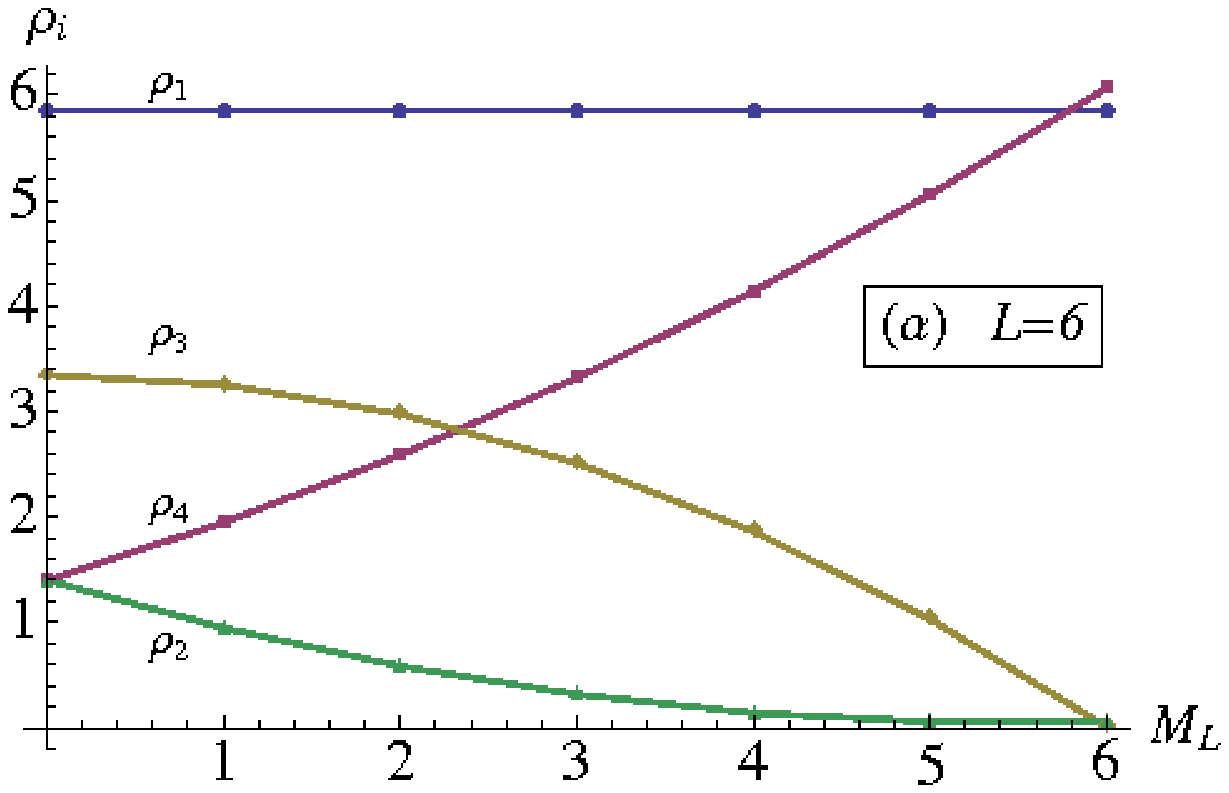} 
   \label{occupation-ml_12_bos}
   \includegraphics[scale=.60]{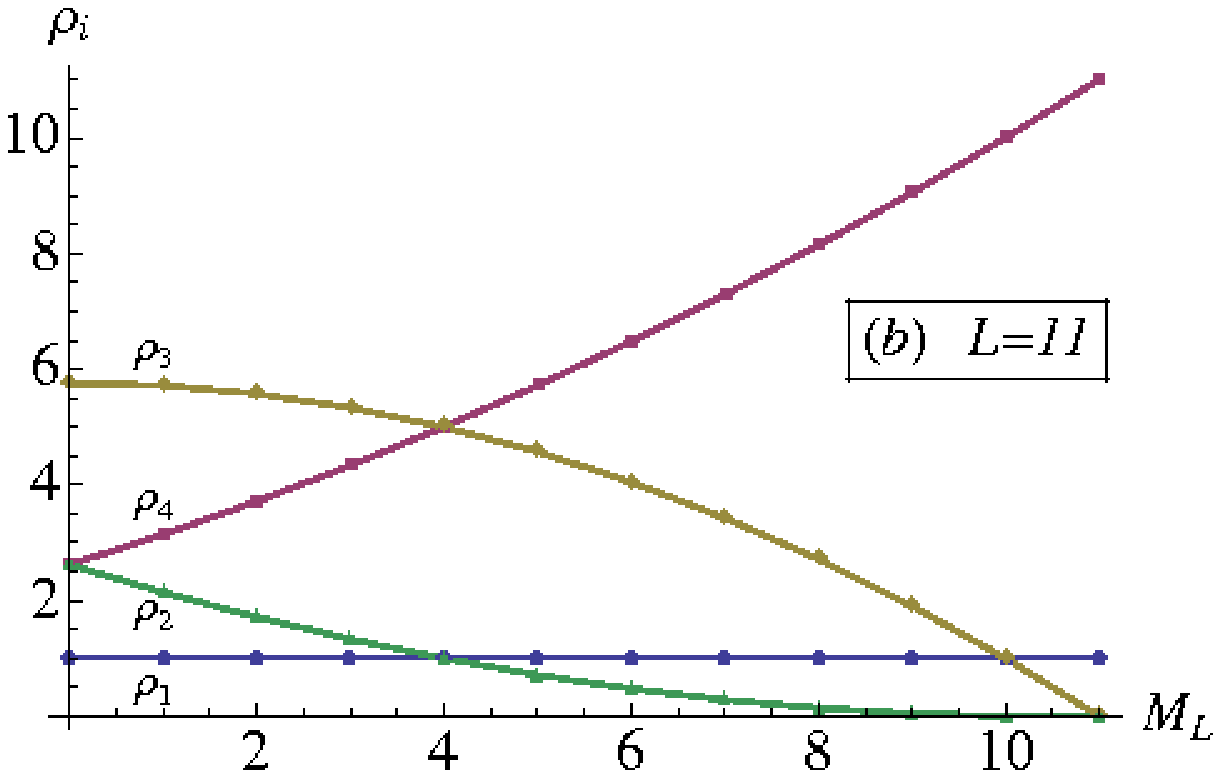}
   \label{variance-ml_12_bos}
   \vspace{-10pt}
 \caption{(Color online) Occupations with respect to $M_L$, for a system of $N=12$ bosons, with interaction strength set to $\lo=-0.842$, in the ground state of angular momentum $(a)$ $L=6$ and $(b)$  $L=11$. The occupations of the three excited natural orbitals differ significanlty when, inside the $L=const.$ subspace, we increase the projection $M_L$ of the angular momentum. The occupation numbers presented here are calculated at the optimal values of $\so, \se$. Both the energy and the optimal $\sigma_i$, $i=0,1$, are invariant under changes of $M_L$. Note that, for different values of $L$, the same pattern on the occupations $\rho_i$ persists, though $\rho_1$ is fixed at different values, according to $\rho_1\simeq N-L$ [see Eq. (\ref{LMBdep})]. All quantities are dimensionless.}
 \label{varoccu-ml_12_bos}
\end{figure}

Every state $\psil$ with definite angular momentum $L$ is $(2L+1)$-fold energetically degenerate, due to the quantum number $M_L$. This means that the energy landscape of a state would not feel any change in $M_L$. Recall from Eq. (\ref{Lzeiv}) that the eigenvalue of $\hat L_z$ of a permanent $\phir$ is simply $M_L=n_4-n_2$. Similarly it can be shown that, for a general state $|\Psi\rangle$, $M_L=\rho_4-\rho_2$ holds. Non surprisingly, this suggests that the occupations $\rho_1, \rho_3$, i.e., the occupations of the two $m_l=0$ orbitals, do not contribute to the z-projection of the total angular momentum $\hat L$. However, as the quantum number $M_L$ of a state with a given $L$ varies, only the occupation $\rho_1$ remains unchanged, while $\rho_3$ varies accordingly to keep the total number of particles fixed, i.e.,
$\rho_3=N-\rho_2-\rho_4-\rho_1=N-M_L-2\rho_4-const.$. This behaviour is depicted in Fig. \ref{varoccu-ml_12_bos} for a system of $N=12$ bosons in its ground state, for $L=6$ on the first and $L=11$ on the second panel. On panel \ref{varoccu-ml_12_bos}$(a)$, at point $M_L=1$, the first from above line (blue) corresponds to the occupation $\rho_1$, the second (yellow) to $\rho_3$, the third (purple) to $\rho_4$ and the fourth one (green) to $\rho_2$. On panel \ref{varoccu-ml_12_bos}$(b)$ the sequence is $\rho_3, \rho_4, \rho_2, \rho_1$, with the same coloring. The occupation numbers presented here are calculated at the optimal values of $\so, \se$, that minimize the total energy of the system. Both the energy and the optimal $\sigma_i,~i=0,1$, are invariant under changes of $M_L$. By comparing the two panels we see that the same pattern on the changes of the occupations is repeated, with $\rho_1$ fixed at different values; at $L=6$, $\re \simeq 6$ while at $L=11$, $\re \simeq 1$. We infer that the behaviour of the occupations against $M_L$ is a general feature, independent of $L, N$.

\begin{figure}
\centering
  \label{frag_1}\includegraphics[scale=.50]{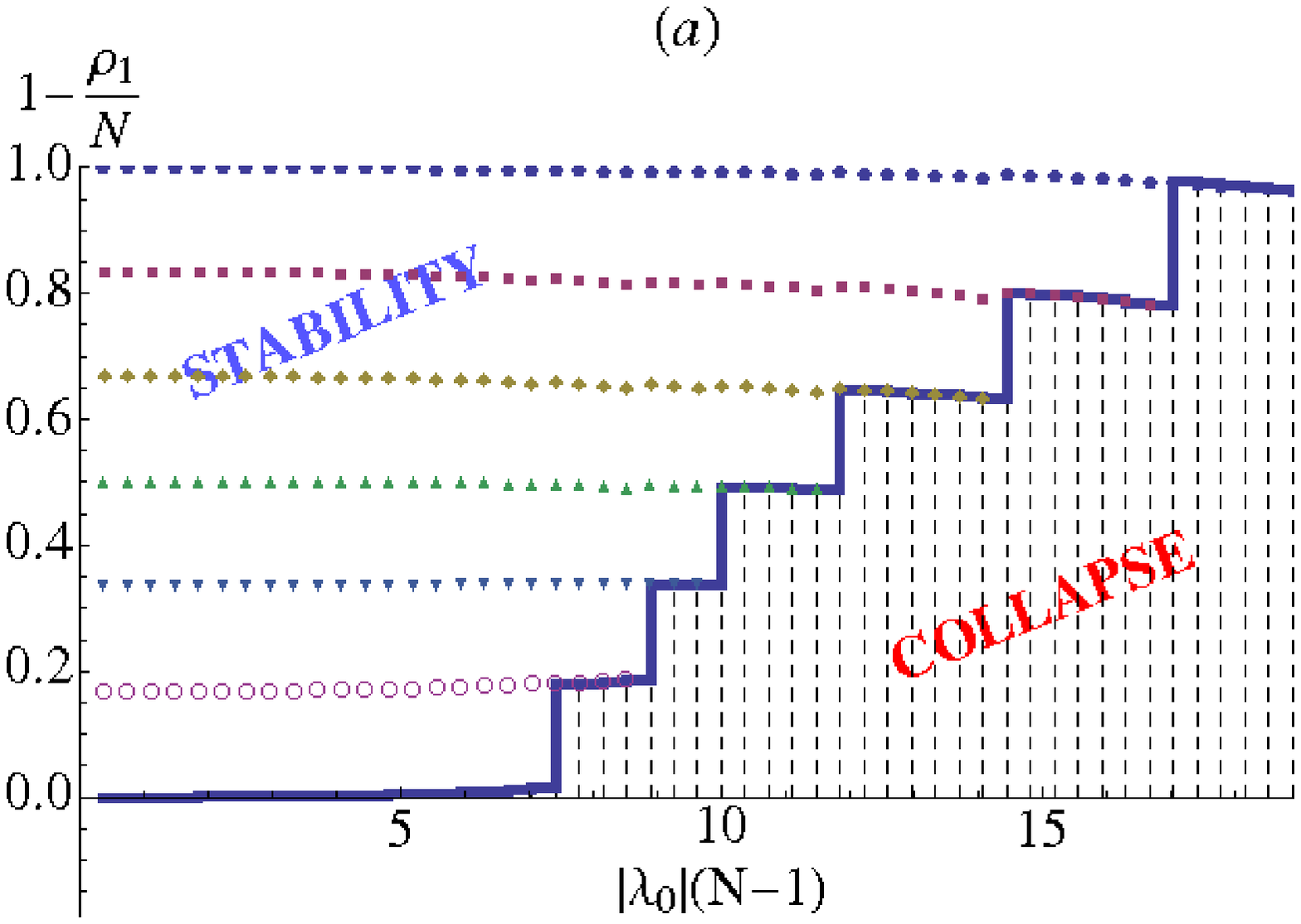} 
  \label{frag_2}\includegraphics[scale=.50]{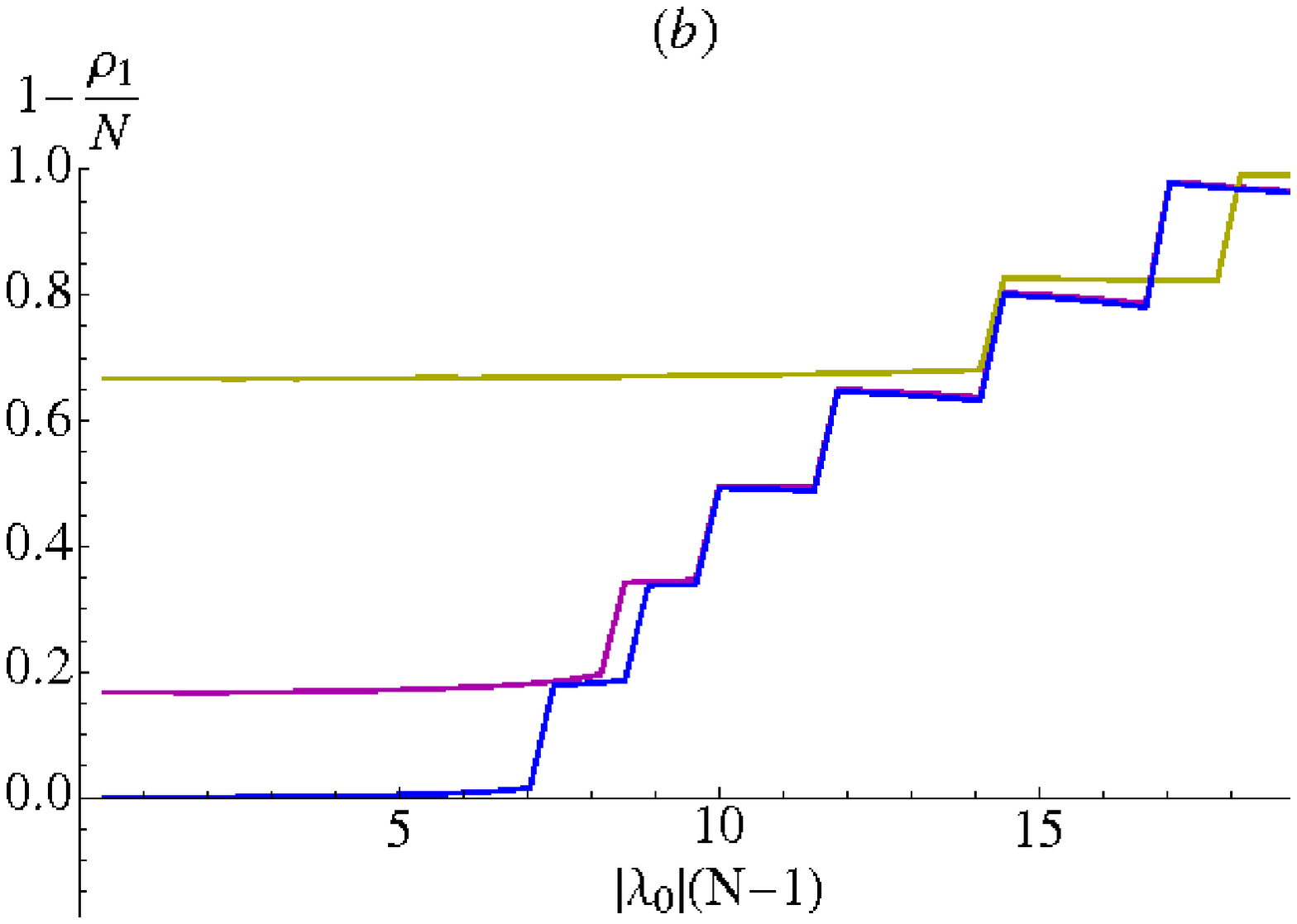} \label{depletion for diff L}
  \caption{(Color online) s-depletion, $d_1=1-\frac{\rho_1}{N} $, varying with the (absolute value of the) interaction strength for a system of $N=12$ bosons. In panel $(a)$ the dotted lines correspond to the ground and the six excited states of the $L=0$ block of the Hamiltonian. The solid line marks the lowest-in-energy metastable (yrast) state that was found at each $\lo$ point. In panel $(b)$, shown are three curves corresponding to the MB (yrast) states of different angular momentum $L$; the lowest one with $L=0, M_L=0$ (blue), the middle one with $L=2, M_L=0$ (magenta), and the upper one with $L=8, M_L=0$ (yellow line). For weak interaction strength the ground state is the $|\Psi^{L=0}_{i=1}\rangle$ state with (almost) zero fragmentation. For $\La=|\lo|(N-1)\simeq8$ the lowest-in-energy state that survives the collapse is the fragmented state $|\Psi^{L=0}_{i=2}\rangle$ with $d_1\simeq0.2$. Its energy is very close to that of the \emph{ground state} $|\Psi^{L=2}_{i=1}\rangle$ of the $L=2$ block. Compare also the three states $|\Psi^{L=8}_{i=1}\rangle$, $|\Psi^{L=2}_{i=4}\rangle$ and $|\Psi^{L=0}_{i=5}\rangle$ at point $\La\simeq15$ (see text for more details). All quantities are dimensionless.}
\label{fragmentation_lambda}
\end{figure}

In Fig. \ref{fragmentation_lambda}, for a system of $N=12$, we show how the depletion $d_1$ varies with increasing absolute value of interaction strength. In the left panel the dotted lines correspond to the six excited states, $i=2,\dots7$, of $L=0$. The solid line marks the depletion of the lowest-in-energy metastable state (ground state), at each value of $\lo$, at the optimal $\so, \se$.  The successive `jumps' of this line take place at the critical values $\lo^i$ where the state collapses. Thus the plane of the figure is divided into the right `collapsed half-plane' and the left `metastable half-plane'.
Similar tendencies persist for states of different angular momenta. This is shown in the right panel, where we plot three curves that correspond to MB states of different angular momenta; the lowest one with $L=0$, the middle one $L=2$, and the upper one with $L=8$, all with $M_L=0$. Each curve is the value of the s-depletion of the lowest-in-energy metastable state with specific $L$ against $\La=|\lo|(N-1)$. For low interaction strength ($\La<7$) the ground state is the state with $L=0$  and (almost) zero fragmentation. For larger values of the interaction strength the condensed state cannot support a metastable state anymore. Though, the first excited (and fragmented) state $|\Psi_{i=2}^{L=0}\rangle$ is found to be non-collapsed.

An examination of the s-depletions of the different ground states of the right panel of Fig. \ref{fragmentation_lambda} allows one a comparison of the respective energies; indeed, two states $|\Psi_{i}^L\rangle$ with the same s-depletion are expected to have the same energy.
For example, the s-depletions of the states $|\Psi^{L=0}_{i=2}\rangle$ and $|\Psi^{L=2}_{i=1}\rangle$ (first and second from below lines, respectively) are very close to each other for the whole range of $\lo$ that they exist and their energies $E[|\Psi^{L=0}_{i=2}\rangle]$ and $E[|\Psi^{L=2}_{i=1}\rangle]$ are found to behave accordingly. In fact, those two states belong to a family of states $\{|\Psi^{L=L_k}_{i=i_k}\rangle\}_k$, whose members, defined by:
\begin{equation}
 \label{iaib}
i_{k}+\frac{L_k}{2}=q, ~q\in \mathbb N^*,
\end{equation}
 have, for $\lo=0$, the same energy, i.e.,
\begin{equation}
\label{Edeg}
 E[|\Psi^{L=L_k}_{i=i_k}\rangle] \stackrel{\lo=0}{=} const.
\end{equation}
for all possible $L_k,i_k$. That is, all the states with $L=L_k,i=q-L_k/2$, for some positive $q\in \mathbb N^*$, are degenerate in the absence of interaction. The degeneracy of such a group of states has been already noted in Ref. \cite{Mottelson1999} and subsequent works. However the states considered there are those of $M_L=L$ and hence the description becomes essentially two dimensional. In the case of $\lo<0$ and $L_1<L_2$ Eq. (\ref{Edeg}) transforms to:
\begin{equation}
\label{Edegalmo}
 E[|\Psi^{L_1}_{i_1}\rangle] \stackrel{\lo<0}{<} E[|\Psi^{L_2}_{i_2}\rangle].
\end{equation}
Namely, the decrease in the energy is larger in the state with the lowest angular momentum, when the attraction is switched on. This behaviour can be seen in the comparison of the states of different angular momentum $L$, on the right panel of Fig. \ref{fragmentation_lambda}.

Summarizing, we see that the s-depletion is an informative quantity of the state, as it reveals information about the energy and the angular momentum, that $|\Psi\rangle$ carries. The s-depletion of a particular metastable state remains almost fixed for $\se,\so\gg1$, while it changes rapidly as the system is driven to the collapse region of the surface. The s-depletion does not depend on the angular momentum $M_L$. Among states with different symmetries (quantum numbers) that are energetically degenerate at $\lo=0$, the attractive interaction favours energetically the one with lower $L$, hence smaller $M_L$-degeneracy.

\subsubsection{Variance}

Besides the s-depletion of the condensate, the variances $\tau_i$ and $\tno$, defined in Eqs. (\ref{tau}) and (\ref{taunorm}), give information about both the structure of the stationary states and the dynamical behaviour of them. Although the calculation of time-dependent states are beyond the scope of this work, one can, based on the present results, comment on the expected dynamical stability of the states. In a fully variational time-dependent multi-configurational approach \cite{RoleOfExcited, MCTDH} both the permanents and the expansion coefficients are time-dependent, i.e., $|\Psi(t)\rangle=\sum_k C_k(t) |\Phi_k(t)\rangle$. As shown in Ref. \cite{CIE} the expansion coefficients $C_k$ in $|\Psi\rangle=\sum_i C_i |\Phi_i\rangle$ comprise a Gaussian distribution on their own, of width characterized by variance $\tv$. So, a state with a large value of $\tv$ will include a large number of coefficients $C_k$ in its expansion. For this reason it is expected to be dynamically more unstable than a state with small $\tv$.

\begin{figure}
\centering
\vspace{0pt}
\includegraphics[scale=.65]{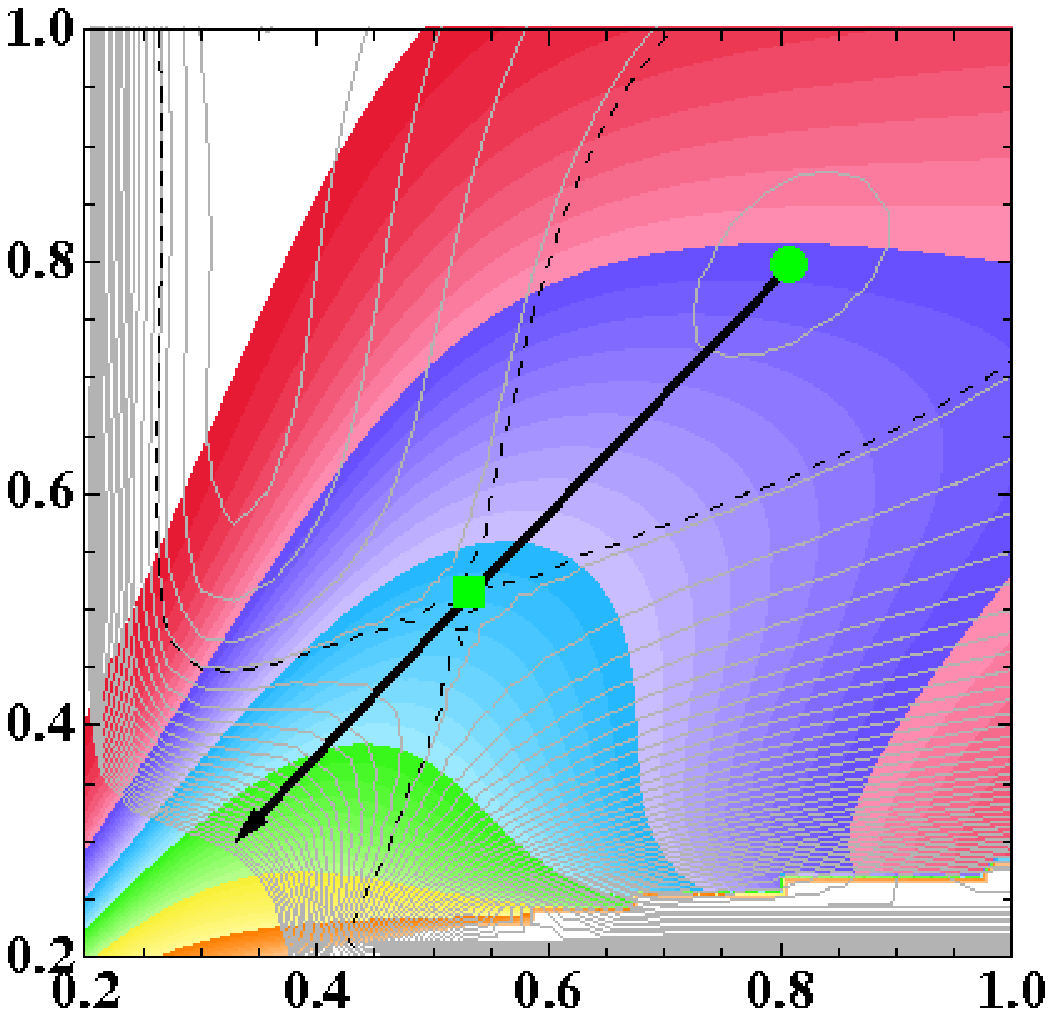}
\hspace{10pt}
\includegraphics[scale=.115]{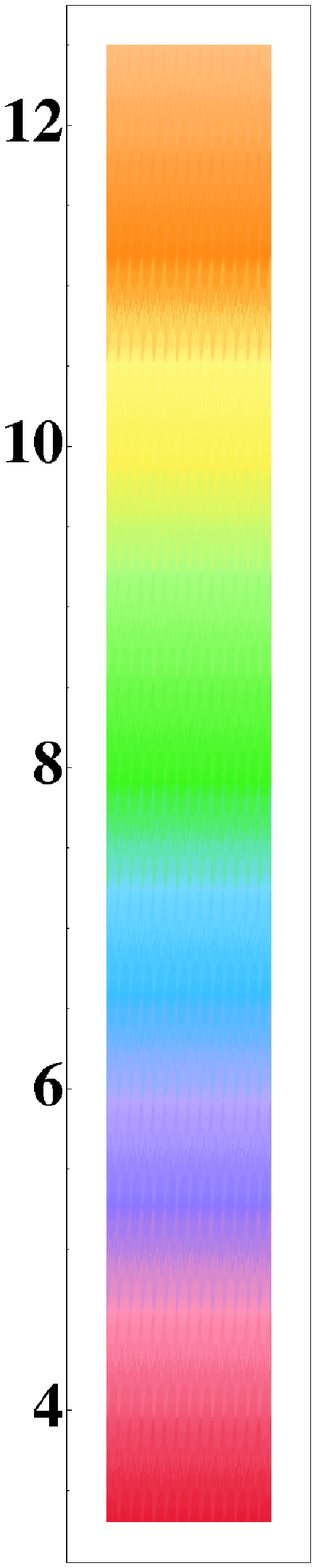}
\vspace{8pt}
\caption{(Color online) Change in variance $\tv$ on the ($\so,\se$) plane for a $N=60$ bosons system, for the same state as in Fig. \ref{Occupations-sig_60}. Plotted are the contours of $\tv=const.$ as well as the contours of constant energy (grey curves). At the minimum of energy (green dot) $\e_0=1.55$, $\so=0.81,\se=0.79$, the variance of the system is $\tv=4.79$. The arrow joins the minimum (dot) and the saddle point (square) of the energy surface, i.e., it depicts the `collapse path'. As the system moves along this `collapse path' the change of the variance grows moderately large (see text for more details). All quantities are dimensionless.}
\label{Variations-sig_60}
\end{figure}

To study the variance of the states $\psil$ we plot in Fig. \ref{Variations-sig_60} the contours of fixed variance $\tv=const.$ on the ($\so, \se$) plane for a system of $N=60$ bosons 
in the ground state with $L=26, M_L=2$. We also draw the `collapse path' (arrow) as defined before, the minimum (dot) and the saddle point (square) of the energy surface as well as the contours of constant energy $\epsilon$ (grey lines). At the minimum of energy, at point $\e_0=1.55$, $\so=0.81,\se=0.79$, the variance of the system is $\tv=4.79$. As the systems moves along the `collapse path' on the energy surface it crosses contours of different variance $\tv$ towards larger values. Since a zero (or almost zero) value of $\tv$ is indicative of a MF state, we see that the system moves, in this way, towards less and less MF states. On the other hand, for large values of the scaling parameters, i.e., $\se, \so\gg 1$, the variance $\tv$ remains practically unchanged.

We have also examined the case of a non collapsed GP ground state of zero angular momentum. For values of parameters $N=12$ and $\lo=-0.5052$ the ground state of the system is the condensed state with $L=0$ and the variance $\tv$, as well as the s-depletion $d_1$, at the optimal $\so,\se$ is almost zero. The same as before scenario is found to hold; in a neighbourhood of the minimum of the energy, in the ($\so,\se$) plane, the variance remains very close to zero but as the system moves over the energy barrier the variance grows larger, i.e., the system moves towards non MF states. The same happens to the s-depletion $d_1$.
Note that, in all cases, the minimum value of $\tv$ and the optimum one (i.e., the value of $\tv$ at the minimum of energy) do not coincide.

\begin{figure}
\centering
\includegraphics[scale=0.59]{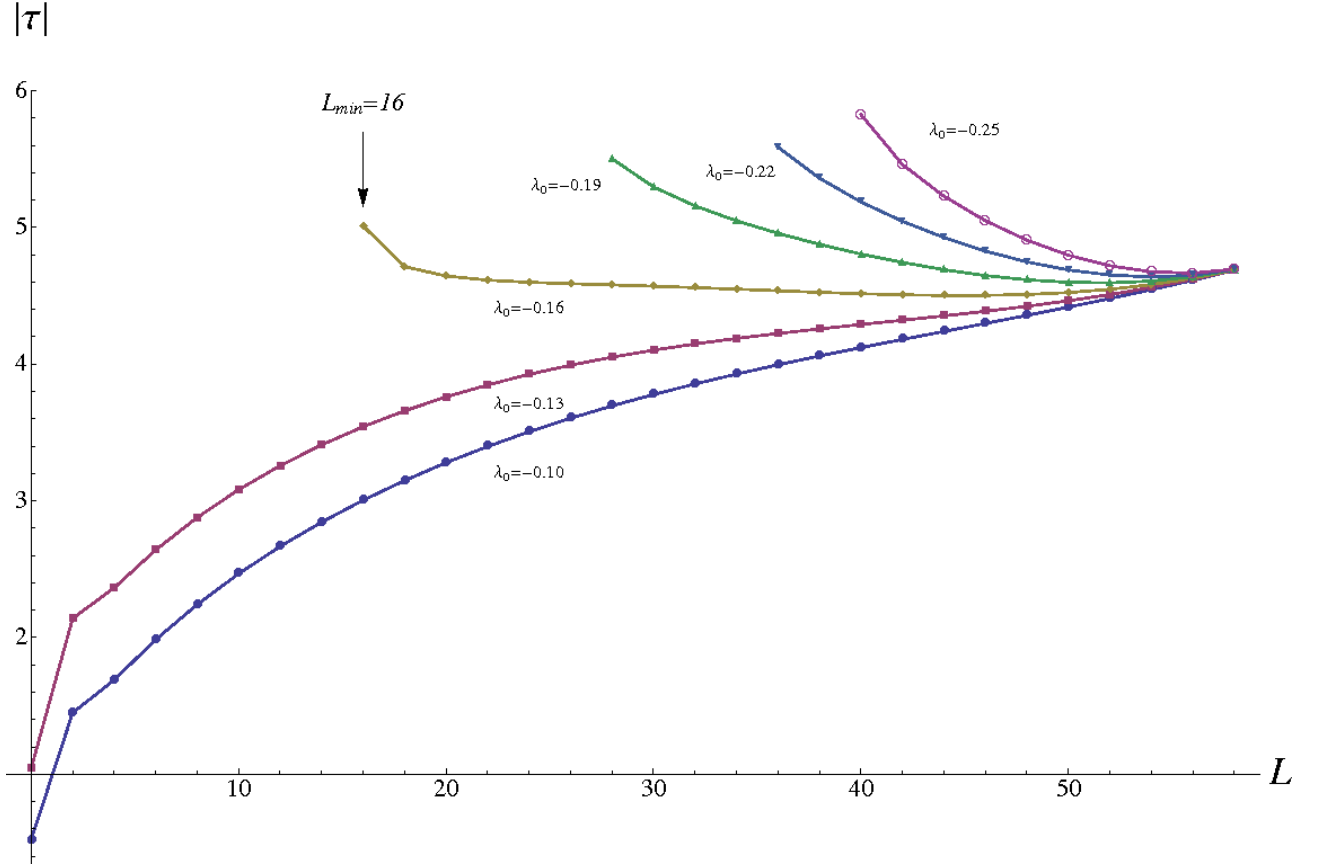}
\caption{(Color online) Change in the variance $\tv$ of ground states $\psile$ against the quantum number $L$, for six different values of the interaction strength $\lo$. The number of particles is $N=60$ and the maximum angular momentum is $L_{max}=N$ (in the diagrams up to $L=58$). The curves shown are for metastable states which do exist. As the value of $\lo$ grows larger the $L$-states, starting from $L=0$ upwards, collapse and hence cease to exist. $L_{min}$ is the minimum value of $L$ that, at each value of $|\lo|$ a metastable ground state of angular momentum $L_{min}$ exists. As an example $L_{min}$ is indicated by an arrow for the $\lo=-0.16$ curve.
For small values of $\lo$, where $L_{min}=0$, the variance of the states increases monotonously with $L$. For larger values of $\lo$ a minimum of $\tv$  appears at some $L>L_{min}$. All quantities are dimensionless.}
\label{variations_l_la}
\end{figure}

In Fig. \ref{variations_l_la} we plot the change in variance $\tv$ of ground states $\psil$, against the quantum number $L$, for six different values of the interaction strength $\lo$. The number of particles is $N=60$ and $L$ varies from $L=0$ to $L=58$ or $0<L/N<0.97$. As we increase the value of $|\lo|$ the $L$-states, starting from $L=0$ upwards, collapse and hence cease to exist. We denote with $L_{min}$ the minimum value of $L$ with which, at a given value of $\lo$, a metastable ground state of angular momentum $L_{min}$ can exist. For small values of $|\lo|$, where $L_{min}=0$, the variation of the states increases monotonously with $L$. For larger values of $|\lo|$ ($\lo\lesssim-0.15$) a minimum in the curve $\tau(L)$ appears, at a point $L>L_{min}>0$. The variances for all different values of the interaction strength meet at one point, as $L\rightarrow N$. Generally we detect two competing tendencies on $\tv$ as $L$ increases; first, since the size of the configuration space $N_p$ drops linearly with $L$ ($N_p=1$ when $L=N$) the number of coefficients in the expansion of Eq.~(\ref{Psi}) decreases with $L$ and so will $\tv$. On the other hand, as $L$ grows larger, the configurations $\overline{\Phi}$ include more basis-functions $\Phi$ in their expansion and hence their variance $\tv_{\overline \Phi}$ increases. The `dominance' of the one or the other tendency seems to be conditioned by the value of the interaction strength $\lo$. However, for large values of $L$, the dependence of $\tv$ on $\lo$ is not significant.

We, next, study the dependence of the variance $\tv$ of the states $\psil$ on the quantum number $M_L$.
We recall that the maximum angular momentum $L_{max}$ that a MB state can possess is, due to the orbital subspace used here, always equal to the total number of particles $N$. The $(2L+1)$ $M_L$-states, of different z-projection of $\hat L$, make every $L$-eigenstate $(2L+1)$-fold degenerate.

\begin{figure}
\centering
 \includegraphics[scale=0.88]{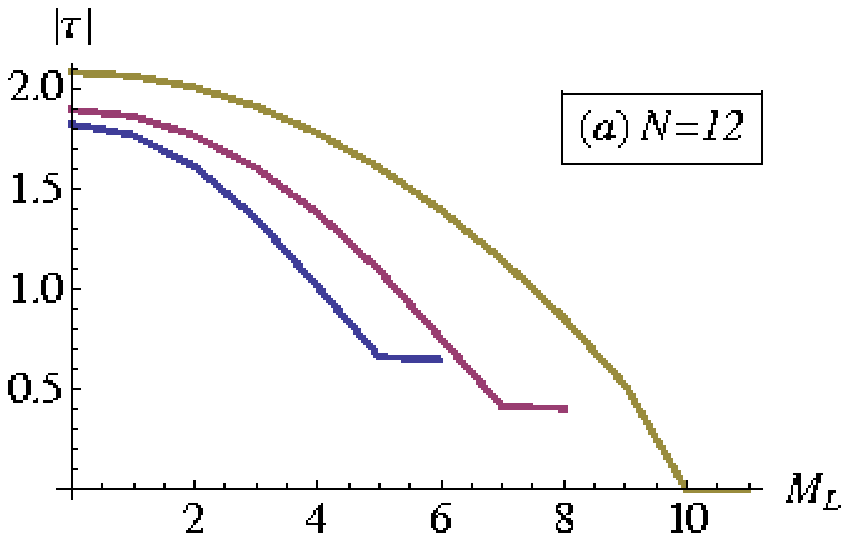}
 \includegraphics[scale=.87]{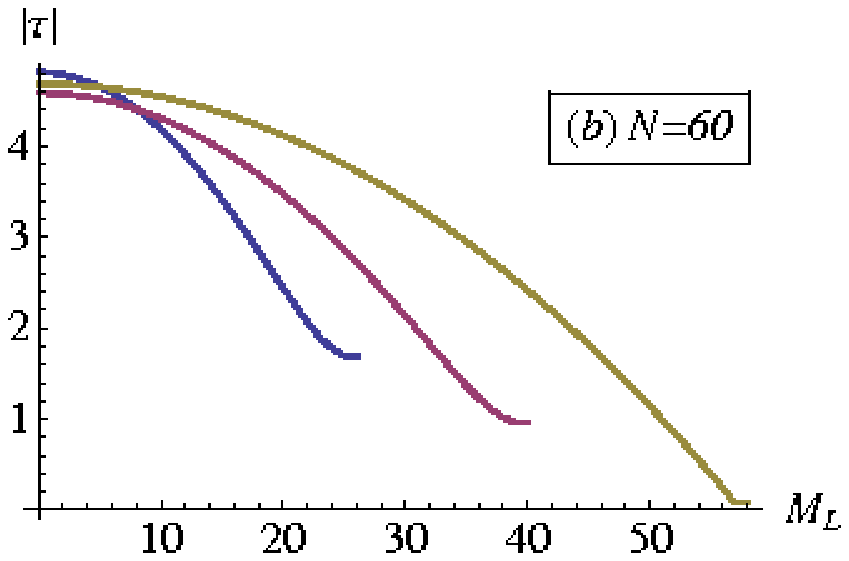}
 \vspace{0pt}
\caption{(Color online) Change in the variance $\tv$ of MB states $\psile$ against the quantum number $M_L$. Precisely, for a system $(a)$ of $N=12$ bosons and interaction strength $\lo=-0.842$ and $(b)$ of $N=60$ bosons and interaction strength $\lo=-0.1684$ we pick three ground states $\psile$ with different values of $L$. On the left panel, at $M_L=4$, the bottom line (blue) corresponds to $L=5$, the medium one (purple) to $L=8$ and the upper (yellow) one to $L=11$. On the right panel, at $M_L=20$, the bottom line (blue) corresponds to $L=20$, the medium one (purple) to $L=40$ and the upper (yellow) one to $L=58$.
In the `edge' of each $L$-block of the Hamiltonian matrix $\mathcal H$, where $M_L=\pm L,\pm(L-1)$, the variance takes always its minimum value. For $M_L=\pm L=\pm N$ the variance $\tv$ is zero and the state is a MF state (see Appendix \ref{appen2b}). The shown decrease of the variance is attributed to the decreasing number of available permanents $\Phi$ that comprise the basis functions $\overline{\Phi}=\mathcal U \Phi$, as the number $M_L$ increases (see text for more details). All quantities are dimensionless.}
\label{Variance_vs_ml-60_Bosons}
\end{figure}

In Fig. \ref{Variance_vs_ml-60_Bosons} we plot the variance $\tv$ as a function of $M_L$ for various states. For systems of $(a)$ $N=12$ and $(b)$ $N=60$ bosons we choose three different ground states $|\Psi_{i=1}^L\rangle$ with $L=5,8,11$ and $L=26,40,58$ (first and second panels, respectively). In the figure, at $M_L=4$ for the left and $M_L=20$ for the right panel, the lowest, middle and upper curves correspond to the lowest, middle and upper values of $L$, respectively (blue, purple and yellow colors).
As the quantum number $M_L$ increases the variance $\tv$ drops, contrary to the fact that the size of the configuration space $N_p$ does not depend on $M_L$ (see Appendix \ref{appen1}). However, the size of the expansion of the basis functions $\overline \Phi$ scales like $(N-|M_L|)^2$ and this results in the decrease of the variance $\tv_{\overline{\Phi}}$ of each of the functions $\overline{\Phi}$, as $M_L$ increases.
In the `edge' of each $L$-block, where $M_L=\pm L,\pm(L-1)$, the variance takes always its minimum value (see also Appendix \ref{appen2b}). If, further, $L=N$ and $M_L=\pm L=\pm N$ the variance $\tv$ is zero, since there is only the permanent $|0,N,0,0\rangle$ (or $|0,0,0,N\rangle$) that contributes to the state $|\Psi\rangle$.

Note that the shown dependence of the variance $\tv$ on $M_L$ is connected to the size of the (truncated) space of one-particle basis functions that we use. In similar calculations over an extended (i.e., less truncated) $\phi$-space, there would be more terms in the expansions of $\overline \Phi$ and the variances shifted to higher values. However the general tendencies, as shown in Figs. \ref{variations_l_la} and \ref{Variance_vs_ml-60_Bosons} are not expected to change.

In this section we have studied the dependence of the variance $\tv$ of a state on the parameters $\so,\se$ and the quantum numbers $L$ and $M_L$. Generally, as the system moves towards collapse (i.e., $\se,\so\rightarrow0$) the variance $\tv$ increases. Moreover, the variance as a function of $L$ can increase monotonously or exhibit a minimum, depending on the value of $\lo$. The variance $\tv$ decreases with increasing $M_L$.

\section{Angular momentum and collapse: Many-Body vs. Mean-Field}
\label{comparison}

As already discussed, any three dimensional attractive condensate is expected to collapse when the product $\La=|\lo|(N-1)$ exceeds a critical value $\La_{cr}$. However, fragmented metastable states can survive the collapse for a much greater value $\La>\La_{cr}$.
In this section we examine the behaviour of MB states $\psil$, as well as these of the MF states $\phir$ of various angular momenta $-$ exact or expectation values $-$ in the onset of collapse. Combining the findings of the previous discussion we show how the angular momentum can stabilize an overcritical condensate. We first discuss the impact of angular momentum on the stability of MB states. We then give an account of the estimated angular momentum within the MF approximation by deriving relevant quantities (expectation value of the angular momentum operator) that will allow us comparisons with the MB results.

\subsection{Many-Body predictions}

In the previous section, we described the structure of MB states that have a definite angular momentum $0\leqslant L\leqslant N$. We showed that, generally, these states are fragmented and, moreover, are non MF states. This suggests that a MB state $\psil$ with definite $L$ can, depending on its s-depletion and the value of $|\lo|$, survive the collapse. Additionally, the condition $[\hat H,\hat L]=0$ necessitates the conservation of the total angular momentum and thus the stability of the state $\psil$. 

Figure \ref{Fragmentation-ALL_BOSONS} summarizes and aggregates the main results of this work. We first focus on the upper connected dotted lines, which are the results for the MB states. For systems of different particle numbers $N=12,20,60$ and 120 (see the legend of the figure for the correspondence to the different colors) we plot the s-depletion $d_1$ versus the quantity $\La=|\lo| (N-1)$. Each plotted point, at each value of $\La$, is the depletion $d_1$ of the ground (yrast) state $\psile$ of some angular momentum $L$ which is still non-collapsed. As the absolute value of the interaction strength increases, the lowest-in-energy states $\psil$ start to collapse. The energies and occupations (depletions) are calculated at the optimal values of the parameters $\so,\se$. As we have already seen in Sec. \ref{fragmentation}, at a given $\lo$, the s-depletion of a MB ground state gives also the angular momentum $L/N$ of this state. Qualitatively, for the ground state of each $L$-block, one can write 
\begin{equation}
\label{LMBdep} 
\frac{L}{N}=1-\frac{\rho_1}{N}+O\left(\tau(\lo)\right) \equiv d_1 + O\left(\tau(\lo)\right),
\end{equation} i.e., the angular momentum of a ground state $\psil$ and the depletion of it differ only to some term $O(\tau)$, that depends on the fluctuation (variance) of that state, which in turn depends on the strength of the interaction. In a non-interacting system the fluctuations are zero and $d_1=\frac{L}{N}$ exactly.

Interpreting the results of Fig. \ref{Fragmentation-ALL_BOSONS} we can say that for any value of the factor $\La$ there will be some $L>0$ such that the (ground) state $\psile$ is metastable. The critical angular momentum $L$ increases monotonously with $\La$. The stability behaviour seems not to depend significantly on the number of bosons in the following sense: for small particle numbers the curves of Fig. \ref{Fragmentation-ALL_BOSONS} are slightly different, while for $N\gg1$ all curves converge, rendering in such a way the obtained results universal and independent of a particular choice of $\lo$ or $N$.

\subsection{Mean-Field predictions \label{MFpredictions}}

Any MB state $\psil$, as we saw, is an eigenfunction of the operator $\hat L^2$. At the MF level however every state $\phir$ of the system is represented by only one permanent, Eq. (\ref{Phi_BMF}). Hence, with the exceptions of states with $M_L=\pm L,\pm (L-1)$, a MF state $\phir$ is by construction incapable of describing eigenstates of $\hat L^2$ (see Appendix \ref{appen2b} for the possible MF states that are eigenstates of the total angular momentum operator). This incapability comprises a major difference between the two descriptions.
Within the multi-orbital BMF \cite{BestMeanField} theory the occupations $n_i$ of each orbital of the ground state are varied to extremize the energy functional of this state. However, in the description of excited states \cite{cederbaum:040402} they serve as parameters that are externally determined. In such a way one is free to choose the values for the set of the occupations $\{n_1,n_2,n_3,n_4\}$ or $\{n_2,n_3\}$  for given depletion $d_1$ and total particle number $N$. So, for example, the choice $n_2=n_3=n_4\neq0$, made in Ref. \cite{cederbaum:040402}, guarantees the sphericallity of the one-particle density [i.e., $\rho(\textbf r)=\rho(r)$], but breaks the  $L$-symmetry of the state. We recall that in the present MB approach the natural occupation numbers, for \emph{all} the ground and excited states, are determined  variationally from the eigenvectors $\mathcal C$ of the optimized Hamiltonian matrix $\mathcal H$, see Eq. (\ref{occupation}). As a result, the rotational symmetries of the system are restored.

So, what is the angular momentum that MF states have? It is a matter of fact that at a MF level one can only speak of expectation values and not exact values/quantum numbers  of $L$. 
It can be shown (Appendix \ref{L-comparison}) that the expectation value $\langle \hat L^2 \rangle$ of the angular momentum of a MF state with equally distributed excited bosons $n_2=n_3=n_4$ is the statistical average (mean) of the exact total angular momentum of the MB states with the same value of depletion $d_1$:
\begin{equation}
\label{LMF}
\tilde L_{MF}=\langle L_{MB}\rangle_{d_1},
\end{equation}
where $\tilde L_{MF}(\tilde L_{MF}+1)=\langle \hat L^2 \rangle$. So, in accordance to its name, the \emph{mean-field} state can provide only the \emph{mean} angular momentum of the corresponding (i.e., same $d_1$) MB states. Furthermore, one can calculate (Appendix \ref{L-comparison}) the average momentum over, first, all the MF states $\phir$ and, second, over all MB states. It turns out that they are connected through:
\begin{equation}
  \langle \tilde L_{MF} \rangle_{all~states} = \frac{N}{\sqrt{5}} = \frac{3}{\sqrt{5}}  \langle L_{MB} \rangle_{all~states}.
\end{equation}
So, the average angular momentum over MF states is found to be $\frac{\sqrt{6}}{2} \simeq 1.22$ times higher than the average one over MB states.

Since, within the MF theory, the states $\phir$ do not possess a definite quantum number $L$ we cannot write any exact correspondence between the depletion $d_1$ of the state and the angular momentum $L$, as we did in the case of MB ground states. Instead, we can use $\tilde L_{MF}$ and relate it to  $d_1$ through:
\begin{equation}
\label{LMFdep}
\frac{\tilde{L}_{MF}}{N}=\frac{2}{3} \left(1-\frac{n_1}{N}\right)=\frac{2}{3} d_1.
\end{equation}
This result is taken in the limit $N\gg 1$  (see Appendix \ref{L-comparison}). Note that it does not depend on the value of $\lo$. This reflects the absence of fluctuations on a MF state, which do depend on the interaction strength $\lo$.

How is $\tilde L_{MF}$ related to the stability of the condensate? Recall first, that a system will survive the collapse if, for a given $\lo$, the number of particles that occupy the s-orbital stays below a critical number $N_{cr}$ \footnote{This should not be restricted to the s-orbital. The system will $-$ also $-$ collapse if the numbers of bosons that reside in the p,d,f\dots orbitals exceed the corresponding critical numbers. However these numbers are quite larger than the critical one of the s-orbital and the collapse of the excited orbitals is therefore not of primary significance.}.
 This however does not forbid the total number $N$ of bosons of the system to be larger than this critical number. Indeed a system can exist in a state with $n_1<N_{cr}$ bosons occupying the s-orbital and $N-n_1$ occupying higher-in-energy orbitals. More precisely, any excitations of bosons to p-orbitals may increase the total energy of the system but will contribute to the total stability of it, since the excited p-bosons `feel' less the interaction energy than the s-bosons. This is the reasoning behind the metastability of fragmented states with an overcritical number of bosons, already demonstrated in Ref.~\cite{cederbaum:040402}. Here we further show that a MF state $|\Phi\rangle$ with non-zero expectation value of angular momentum $L_{MF}>0$ exhibits fragmentation, Eq. (\ref{LMFdep}), which increases the overall stabilility of the system. However, the impact of the angular momentum in the stability of the condensate is overestimated at the MF level. A comparison of Eq. (\ref{LMFdep}) with the corresponding MB one, Eq. (\ref{LMBdep}), convinces us of this claim.

\begin{figure}
\centering
\vspace{-25pt}
\includegraphics[scale=0.44]{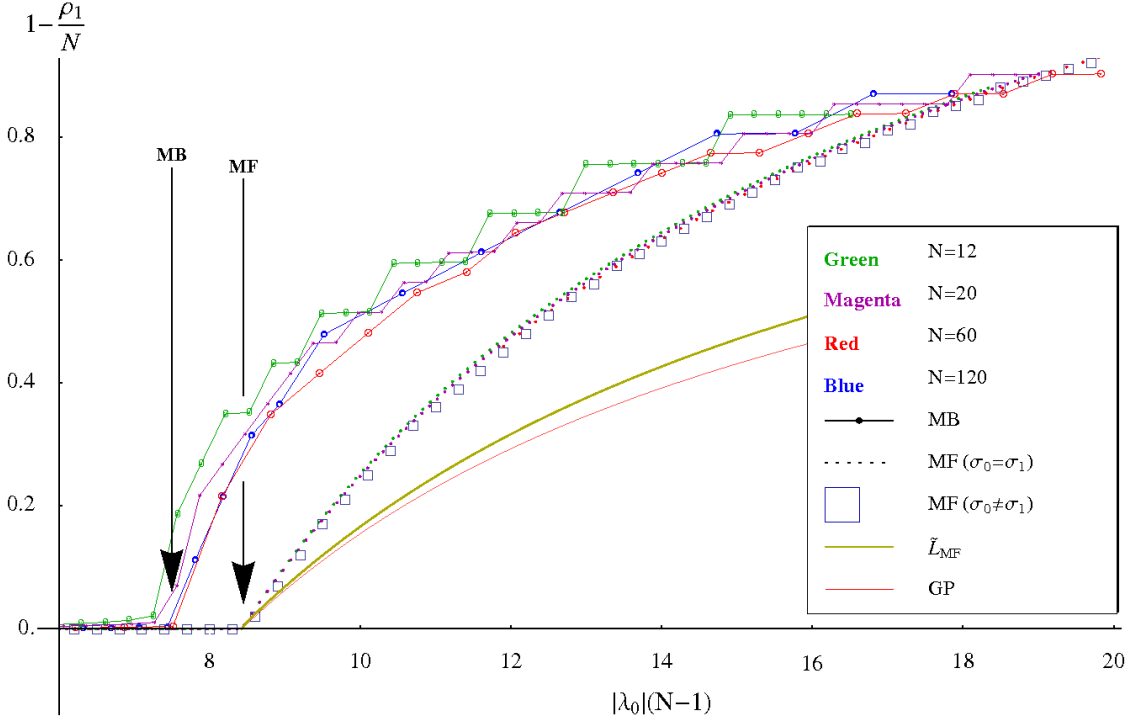}
\vspace{-10pt}
\caption{(Color online) Stability plot for systems of different number of bosons $N=12,20,60,120$. As the interaction strength $|\lo|$ increases the lowest-in-energy states start to collapse; here we plot, at each value of $\lo$, the depletion of the lowest-in-energy state, that is still non-collapsed.
 The connected dotted lines (upper `band' of curves) are the many-body calculations. Every point of the MB plots corresponds to a ground (yrast) state $\psile$ which, unlike the mean-field states, have a definite value of angular momentum $L$. The angular momentum $L/N$ and the depletion of the ground states $\psile$ are almost equal, depending on the fluctuations of the state.
 The dotted unconnected lines are the critical s-depletions, as estimated from MF theory; the calculations here are done for the permanents $|n_1,n,n,n\rangle$, built over four orbitals, with equal occupations of the p-orbitals.
 The second lowest continuous line (yellow) determines the expectation value of the angular momentum $\tilde L_{MF}/N$, over (generally fragmented) MF states, which is given by the depletion $\frac{2}{3} (1-\rho_1/N)$, here $\rho_1\equiv n_1$.
 The lowest continuous line (red) on the diagram depicts the maximum number of bosons $\ncgp$ that can be loaded in a Gross-Pitaevskii condensate (marked as GP in the graph)  without collapsing. For this curve one has to identify the axis $1-\frac{\rho_1}{N}$ with $1-\frac{\ncgp}{N}$, i.e., $\rho_1$ plays the role of the critical GP particle number at a given $|\lo|(N-1)$.
 The difference in the estimation of the factor $\La_{cr}=|\la_{0,cr}| (N-1)$ where the $L=0$ ground state collapses is evident; the overestimated value of $\La_{cr}$ from the GP approach is larger than the MB one and puts the latter closer to the experimentally measured one \cite{Controlled_Collapse}. The depicted stability behaviour does not significantly depend on the particle number $N$, making this behaviour universal (see text for more details). All quantities are dimensionless.}
\label{Fragmentation-ALL_BOSONS}
\end{figure}

To allow a better comparison to the MB results of Fig. \ref{Fragmentation-ALL_BOSONS} we plot on the same graph the data obtained from the MF states (second group of dotted unconnected lines in Fig. \ref{Fragmentation-ALL_BOSONS}). More precisely, for systems of $N=12,20$ and $N=60$ bosons, we plot at each value of $|\lo|(N-1)$  the s-depletion $d_1=1-\rho_1/N$, with $\rho_1$ now given by the critical number of particles $N_{cr}$ (i.e., maximum number of particles so that the state $\phir$ is not collapsed) calculated from the relation:
\begin{equation}
\label{Ncrbmf}
N_{cr}=N_{cr}^{GP} \frac{(-64\cdot5^{3/4}-128\cdot5^{3/4} \frac{n}{N}+300 \Lambda_0-375 \Lambda_0 \frac{n}{N}) }{\left[-4+13 (\frac{n}{N})^2\right](16\cdot5^{3/4}-75 \Lambda_0)},
\end{equation}
at $\so=\se$, with $\Lambda_0=\frac{\lo}{4\pi} \left(\frac{2}{\pi}\right)^{1/2} \frac{4}{3}$, $N_{cr}^{GP}=1-\left(\frac{1}{5}\right)^{1/4} \frac{16}{15} \frac{1}{\Lambda_0}$ and $n=n_2=n_3=n_4$ the occupations of the p-orbitals. Equation (\ref{Ncrbmf}) in the limit $N\gg1$ gives back Eq. (7) of Ref. \cite{cederbaum:040402}. Also, the numerical MF calculations for the critical numbers of a $N=120$ bosons system, without using the assumption $\so=\se$, are presented in Fig. \ref{Fragmentation-ALL_BOSONS} with the `boxed' line (blue). 
The second from below continuous line (dark yellow) determines the angular momentum expectation value $\tilde L_{MF}/N$ [Eqs. (\ref{LMFdep}) and (\ref{Ltil})], over MF states.  The lowest continuous line (red) of Fig. \ref{Fragmentation-ALL_BOSONS} is the calculations from the Gross-Pitaevskii theory. In this case one has to identify $1-\frac{\rho_1}{N}$ with $1-\frac{\ncgp}{N}$, where $\ncgp$ is the maximum number of bosons that, for a given $\lo$, can be loaded in a GP state without collapse. Here we use $N=60$ bosons. The total particle number $\ncgp$ is, of course, the number of s-bosons of the system. Obviously this critical number is decreased, as we move to the right of the x-axis of the diagram and hence we call this curve the `critical GP'.

The `bands' of MF and MB states depicted in Fig. \ref{Fragmentation-ALL_BOSONS} substantially deviate one from each other at small and moderately larger values of $\La$. This is nicely manifested in the difference between the MF and MB predictions of the collapse of the $L=0$ ground state. The collapse of the MB state appears to happen at a smaller value of the product $\La$ than the one that the MF theory estimates. This reflects the overestimation of the impact of the angular momentum within the MF and puts the MB prediction closer to the experimentally measured values of $\La$ (see Ref. \cite{Controlled_Collapse} and also the discussion in Refs. \cite{Mean-Field_Analysis_of_collapsing, Collapsing_BEC_beyond_GP, BEC-Collapse_comparison, Das2009} about the discrepancies between MF predictions and experimental values of the critical numbers and the collapse times).

We see that the form of the curves for the s-depletion of the MF states seems not to be affected from the number $N$ of the particles of the system. The various plotted MF curves for different $N$, like the MB ones, tend to converge for $N\gg 1$, making thus the described stability behaviour a universal and independent of $N$ phenomenon. For the MF case convergence has been noticed already for $N\sim 10^2$ bosons. Note, though that, unlike the MB states, the MF ones with $L=0$ collapse all \emph{at the same} critical value $\La_{cr}=|\lambda_{0,cr}| (N-1)$, regardless of the total number $N$ of bosons.
We also see in Fig. \ref{Fragmentation-ALL_BOSONS} a divergence of the angular momentum $\tilde L_{MF}$ (dark yellow line) from the s-depletion of the MF states (dotted lines); this is exactly the relation of the two quantities, that Eq. (\ref{LMFdep}) provides. The `critical GP' curve significantly diverges from both the multi-orbital MF and the MB predictions.

Conclusively, we presented a way, Eq. (\ref{LMF}), to connect the angular momenta of a MF state of the form $|n_1,n,n,n\rangle$ to that of the MB states with the same depletion $d_1$. A non-zero angular momentum will result in a fragmented condensate [Eq. (\ref{LMFdep})] which in turn will render the system more stable, with respect to the parameter $\La$. Those results are in agreement with the MB ones of the previous section.

\section{Summary and Conclusions  \label{Outlook}}
In this work we constructed many-body states with definite angular momentum quantum numbers $L$ and $M_L$, for systems of $N$ isotropically trapped bosons in three dimensions, interacting via an attractive two-body potential. These many-body states are written as an expansion (configuration interaction expansion) over orthogonal many-body basis functions (permanents). We represented the Hamiltonian and angular momentum operators as matrices on this basis and we looked for the states that simultaneously diagonalize them. 
In this representation the Hamiltonian has a block-diagonal form, with each block consisting of many-body states, with the same eigenenvalue of angular momentum. The one-body basis functions that we used are the wave functions (s- and p-orbitals) that solve exactly the linear (non-interacting) problem, each scaled under a parameter $\s_i$, which we determined variationally. The rotational symmetries as well as symmetries under spatial inversion that the one-body basis functions possess are also present in the many-body states and reduce significantly the size of the configuration space. Due to the truncated one-particle basis set, the total angular momentum is restricted to $0\leqslant L\leqslant N$.  To our knowledge this is the first time that a fully three-dimensional Bose gas in isotropic trapping potential is studied, with the many-body wave function of the system expressly written as an eigenfunction of both total angular momentum operators $\hat L^2$ and $\hat L_z$, for $L\geq0$.

 For a value of the parameter $\La=|\lo| (N-1)$ such that the $L=0$ ground state of the system is collapsed, we have plotted the energy per particle $\e(\so,\se)$ of the ground and the excited many-body states, as a function of the parameters $\so,\se$. We have shown that metastable excited states of the same angular momentum can exist. Furthermore, for the same system, we demostrated the existence of metastable ground states with angular momentum $L>0$ that can survive the collapse. These states would also collapse, if the (absolute value of the) interaction strength is further increased. The examination of the above states, in terms of the natural orbital analysis, revealed that the states are fragmented, with a substantial number of particles being excited to the p-orbitals.

 We discussed why the s-depletion of a many-body state $|\Psi\rangle$ bears information about the energy and the angular momentum of $|\Psi\rangle$. We found that the s-depletion of a metastable state remains practically fixed for $\so, \se \gg 1$, while it changes rapidly as the system is driven to collapse. We have shown also that the z-projection of the angular momentum $M_L$ does not affect the occupation of the first natural orbital.

We have studied the dependence of the variance $\tv$ of a state on the parameters $\so,\se$ and the quantum numbers $L$ and $M_L$. We saw that along the `collapse path' the variance increases. The variance as a function of $L$ $-$ depending on the value of $\lo$ $-$ can increase monotonously or exhibit a minimum. We also found that as the quantum number $M_L$ increases the variance $\tv$ decreases.

 To further investigate the impact of the angular momentum on the stability of the system, we plotted the critical s-depletion of the metastable ground states $\psile$ (yrast states) as a function of the quantity $\La=|\lo|(N-1)$. We showed the connection of the s-depletion to the critical angular momentum $L$, in both the mean-field and the many-body cases. We demonstrated that for any value of the factor $\La$ there is some angular momentum $L>0$ such that the (ground) state $\psile$ is metastable. The critical angular momentum $L$ increases monotonously with $\La$ and this behaviour is found to be independent of the particle number $N$, as long as $N\gg 1$. We derived analytical relations for the expectation value of the angular momentum of a mean-field state, with equally distributed excited bosons, which allowed us to compare it with the corresponding results from the many-body approach. We have further demonstrated that the angular momentum of this mean-field state equals the average angular momentum of many-body states, with the same s-depletion.

Conclusively, we can say that for any particle number $N$ and interaction strength $\lo$ of an attractive condensate, there is some well defined quantum number $L$ of the many-body angular momentum operator $\hat L^2$ such that the \emph{ground state} of this system is metastable, i.e., exhibits a clear minimum in energy as a function of the shapes of the orbitals. Moreover, since the total angular momentum of the system is conserved, once the system is prepared in a ground state with $L>0$ it can survive the collapse, and that for a particle number/interaction strength much beyond the corresponding ones of the $L=0$ ground state. We hope that our results will stimulate experimental research.

\begin{acknowledgments}
Financial support by the HGSFP/LGFG and DFG is acknowledged.
\end{acknowledgments}

\appendix \label{appendix}

\section{Size of Fock Space \label{appen1}}

The total number of the $N$-body basis functions (permanents) that can be constructed over a basis of $M$ one-particle wave functions of Eq. (\ref{orbitals}) is \cite{CIE}:
\begin{equation}
N_p =  \left( {\begin{array}{*{20}c} M+N-1 \\ N \\ \end{array}} \right) = \frac{{(M+N-1)!}}{{N!\left( {M-1} \right)!}},
\end{equation}
which for $M=4$ becomes
\begin{equation}
 N_p = \frac{1}{6} (N+1)(N+2)(N+3)\simeq \frac{N^3}{6}.
\end{equation}
Using the symmetries of the system we can reduce significantly this number and hence the complexity of the problem. Without loss of generality we assume that the particle number $N$ and the quantum number $M_L$ are even integers.
 
\paragraph{Total angular momentum $\hat L_z$:} Since $[\hat L_z,\hat H]=0$ the state $|\Psi\rangle=\sum_i^{N_p} C_i |\Phi_i\rangle$ can be chosen to be a common eigenfunction of the two operators. This transforms $\mathcal H$ to a block diagonal form, with every block consisting of states of distinct $M_L$.
The number of states $|\Psi\rangle$ in a block with some $M_L$ is 
\begin{equation}
N_p=\frac{(N+2-|M_L|)^2}{4}\lesssim \frac{N^2}{4}.
\end{equation}
\paragraph{Parity $\hat \Pi \Psi(\r)=\Psi(-\r)$:} Similarly, $[\hat \Pi,\hat H]=0$ and $\mathcal H$ block diagonalizes into two blocks, each with distinct parity $\Pi=+1$ or $\Pi=-1$.
The number of states $|\Psi\rangle$ in the block with $\Pi=1$ is 
\begin{equation}
 N_p=\frac{\left[N + 4 + 2 M_L - 3 M_L H (M_L)\right] \left[N + 2 - M_L H (M_L)\right]}{8} \lesssim \frac{N^2}{8},
\end{equation}
where $H(x)$ is the unit-step function.
\paragraph{Total angular momentum $\hat L^2$:} Last, the commutator $[\hat L^2,\hat H]=0$, diagonalizes the matrix $\mathcal H$ into blocks of states that have definite angular momentum quantum number $L$.
The number of states $|\Psi^L\rangle$ in the block with some $L$ is 
\begin{equation}
 \label{size_of_block}
N_p= \frac{N-L+2}{2}\lesssim \frac{N}{2},
\end{equation}
where $L$ is the quantum number of $\hat L^2$ and here it is assumed to be an even number. In case $L$ is odd Eq. (\ref{size_of_block}) should read: $N_p=(N-L+1)/2$. Note that these relations hold for any $M_L$.

\section{Angular momentum in Many-Body and Mean-Field theories \label{appen2}}
\subsection{Many-body eigenstates of the total angular momentum operator $\hat L^2$ \label{appen2a}}

We now return to the question of explicitly finding the eigenstates of the operator $\hat L^2$, as discussed in Sec. \ref{Angular}.

A general permanent $\phir=|\vec n\rangle$, representing a system of a total number of bosons $N=n_1+n_2+n_3+n_4$ and z-projection of the angular momentum $M_L=n_2-n_4$, takes on the form:
\begin{equation}
\label{evenperm}
 |\vec n\rangle =|n_1,n_2,n_3,n_4\rangle= |N-2n_2-n_3+M_L,n_2,n_3,n_2-M_L\rangle,
\end{equation}
where $n_2,n_3$ are integers, such that $M_L H(M_L)\leqslant n_2\leqslant (N+M_L)/2$, $0\leqslant n_3 \leqslant N-2n_2+M_L$, where $H(x)$ is the unit-step function. An expansion $|\Psi\rangle$ over these (orthogonal) permanents $\phir$ is:
\begin{equation}
\label{PsiA}
|\Psi\rangle = \sum_{n_2,n_3} C_{n_2,n_3}|\vec n\rangle
\end{equation}
where $n_2, n_3$ run over all possible permanents of Eq. (\ref{evenperm}). Acting operator Eq. (\ref{L2}) on Eq. (\ref{PsiA}) we get:
\begin{eqnarray}
 \label{L2PsiA}
\hat L^2|\Psi\rangle=\hat L^2 \sum_{n_2, n_3} C_{n_2,n_3} |\vec n\rangle =\Lambda \sum_{n_2, n_3} C_{n_2,n_3} |\vec n\rangle ,
\end{eqnarray}
or
\begin{equation}
 \label{L2PsiA2}
\Lambda\sum_{n_2, n_3} C_{n_2,n_3} |\vec n\rangle=\sum_{n_2, n_3} C_{n_2, n_3} {\Big(}A(n_2, n_3)|\vec n\rangle+B(n_2, n_3) |\vec n+2\rangle+\Gamma(n_2, n_3)|\vec n-2\rangle{\Big)},
\end{equation}
where $\Lambda=L(L+1)$ are the eigenvalues of $\hat L^2$, $|\vec n+2\rangle=|n_1, n_2-1, n_3+2, n_4-1\rangle$ and $|\vec n-2\rangle=|n_1, n_2+1, n_3-2,n_4+1\rangle$, i.e., they are the double `excitations' of the permanent $|\vec n\rangle$. 
The functions A,B,$\Gamma$  are:
\begin{equation}
 \label{ABG}
\begin{gathered}
 A(n_2,n_3)=n_2(n_3+1)+n_3(n_4+1)+n_3(n_2+1)+n_4(n_3+1)+(n_2-n_4)^2, \\
B(n_2,n_3)=2\left[n_2 n_4(n_3+1)(n_3+2)\right]^{1/2}, \\
\Gamma(n_2,n_3)=2\left[n_3(n_3-1)(n_2+1)(n_4+1)\right]^{1/2}.
\end{gathered}
\end{equation}

The problem is focused in calculating the coefficients $C_{n_2,n_3}$ such that Eq. (\ref{L2PsiA}) is fulfilled. 
We will show how one can reduce this equation to a simpler form. 
By multiplying Eq. (\ref{L2PsiA2}) with $\langle \vec n|$ and using orthogonality of permanents and the relation
\begin{equation}
\label{Gammabeta}
\Gamma(n_2-k,n_3+2k)=B(n_2-k+1,n_3+2k-2),~ k\in \mathbb N, 
\end{equation}
we obtain:
\begin{equation}
\label{system}
\Lambda C_{n_2,n_3} = A(n_2,n_3) C_{n_2,n_3}+\Gamma (n_2,n_3) C_{n_2+1,n_3-2}+B(n_2,n_3) C_{n_2-1,n_3+2} .
\end{equation}
This is a homogeneous, second order recurrence (or difference) equation of the two independent variables $n_2, n_3$, with known non-constant coefficients.

In the above equations there are two free parameters $n_2,n_3$ which are varied independently and these are also the independent variables of Eq. (\ref{system}). To reduce the dimensionality of the problem one can proceed by switching the representation of the permanents and their coefficients. Precisely, we can use a simpler representation for indexing the vectors $\phir$ in the expansion of $|\Psi\rangle$. Noticing that the action of the operator $\hat L^2$ on a state of Eq. (\ref{PsiA}) involves only permanents of the form 
\begin{equation}
 \label{evenpermi}
|\vec n_i\rangle=|n_1,\a-i,\b+2i,\a-i+M_L\rangle,
\end{equation}
where $\a,\b\in \mathbb N$ and $-(N+M_L)/2\leqslant i \leqslant -M_L H(M_L)$, we can work with permanents of the above type only, for fixed $\a,\b$. In fact the action of $\hat L^2$ partitions the configuration space into invariant subspaces, with permanents of the form of Eq. (\ref{evenpermi}). Permanents with $\a\neq \a' \text{ or } \b\neq\b'$ will not contribute to the same eigenstate $|\Psi\rangle$.
This allows us to move from the \emph{two-parametric} $\{n_2,n_3\}-$ to the \emph{one-parametric} $\{i\}-$ representation. We write now again Eqs. (\ref{PsiA})-(\ref{system}) in the new representation. 

A general state becomes:
\begin{equation}
\label{PsiAi}
|\Psi\rangle = \sum_{i} C_{i} |\vec n_i\rangle,
\end{equation}
where $i$ runs again over all permanents (\ref{evenpermi}). Similarly, acting operator Eq. (\ref{L2}) on Eq. (\ref{PsiAi}) we get:
\begin{eqnarray}
 \label{L2PsiAi}
\hat L^2|\Psi\rangle=\hat L^2 \sum_i C_i |\vec n_i\rangle = \Lambda \sum_i C_i |\vec n_i\rangle~~~~ \\ 
=\sum_i C_i {\Big (}A_i |\vec n_i\rangle+B_i |\vec n_i+2\rangle+\Gamma_i |\vec n_i-2\rangle{\Big )},
\end{eqnarray}
with $A_i=A(\a-i,\b+2i), B_i=B(\a-i,\b+2i)$ and $\Gamma_i=\Gamma(\a-i,\b+2i)$.
Equations (\ref{Gammabeta}) and (\ref{system}) become:
\begin{eqnarray}
 \label{Gammabetai}
\Gamma_{i+2k}=B_{i+2(k-1)},~
\end{eqnarray}
and
\begin{equation}
 \label{systemi}
 (A_i-\Lambda)C_i+\Gamma_i C_{i-1}+B_i C_{i+1}=0,
\end{equation}
respectively.
The above is a homogeneous second-order recurrence (difference) equation of one independent variable [cf. Eq. (\ref{system})]. 

For some choices of the parameters $\a,\b,\Lambda,M_L$ Eq.~(\ref{systemi}) can be easily solved. In particular, for $\a=0, \b=N$ and $M_L=0$ we obtain:
\begin{eqnarray}
 \label{sysresults1}
\Lambda=0~(L=0), &~~ C_i=\frac{(-1)^{2N+1-i} 2^{N+2-2i}\Gamma(\frac{N}{2}+2-i) \Gamma(\frac{N+1}{2}) }{\Gamma(i-1/2)} C_0,  \\
 \label{sysresults2}
\Lambda=2~(L=1), &~~ C_i= \frac{(-1)^{2N+1-i} 2^{N+3-2i}N \Gamma(\frac{N}{2}+2-i) \Gamma(\frac{N+1}{2}) }{(6i-N-6)\Gamma(i-1/2)} C_0,  \\
 \label{sysresults3}
\Lambda=6~(L=2), &~~ C_i=\frac{(-1)^{2N+1-i} 2^{N+5-2i} (N-2) N \Gamma(\frac{N}{2}+2-i) \Gamma(\frac{N+1}{2}) }{ (140 i^2 -60 (N+5) i + 3 N (N+18)+160) \Gamma(i-1/2)} C_0,
\end{eqnarray}
where $C_0$ is to be determined from the normalization condition $\sum_i |C_i|^2=1$ and $-N/2\leqslant i \leqslant0$. Equations (\ref{sysresults1})-(\ref{sysresults3}) give three of the states $\overline \Phi^L$, that are eigenstates of $\hat L^2$ and belong to the rotated basis $\{\overline\Phi_i^L\}$ of Sec. \ref{Angular}.

\subsection{Mean-field and average many-body angular momentum \label{appen2b}}
\label{L-comparison}
We show here that the total angular momentum of a mean-field state, with equally distributed excited bosons $n_2=n_3=n_4\equiv n$ is the statistical average of the exact total angular momentum of the many-body states with the same depletion $d_1=1-\frac{\rho_1}{N}$, i.e.,
\begin{equation}
 \tilde{L}_{MF}=\langle L_{MB}\rangle_{d_1}.
\end{equation}
Recall that $0 \leqslant L \leqslant N$. Then, for a total number of $N$ bosons there are $N+1$ blocks ($L$-blocks) of the Hamiltonian matrix $\mathcal H$, each with a distinct value of $L$.
We want to calculate the average angular momentum $\langle L \rangle$, among states $\psil_{n_1}$ with a given natural occupation $\rho_1=n_1$. We assume that in each $L$-block this occupation $n_1$, as we move from the highest-excited state to the ground state, increases like 
\begin{equation}
\begin{cases}
n_1(k)=2k,   & \mbox{if } L\mbox{ is even}\\
n_1(k)=2k+1, & \mbox{if } L\mbox{ is odd} ,
\end{cases}
\end{equation}
where $k \in \mathbb N$ indexes the state $|\Psi^L_{i=k}\rangle$. The above relations hold exactly in the absence of interaction, i.e., $\lo=0$, and in a satisfactory approximation when $\lo\neq 0$. 
Then, as we have numerically verified, each $L$-block with $L\lesssim N-n_1$ contains exactly one state $\psil_{n_1}$ with the desired $n_1$ (or very close to it). Recall that the size of an $L$-block drops linearly with $L$, as in Eq. (\ref{size_of_block}). So there are $N+1-n_1$ $L$-blocks that contain one state $\psil_{n_1}$. The occupation $n_1$, in the case of $\lo=0$, is even in half of the blocks, odd in the other half ones. The total number of states with occupation $n_1$ is:
\begin{equation}
N(n_1) = \sum_{i=0}^{N+1-n_1,2} (2i+1)
\end{equation}
due to the $\hat L_z$ degeneracy. To include only even (or approximately even) occupations $n_1$ we sum on a step of two (the added term `$,2$' in the upper limit of the sum denotes that step).
These states have total angular momentum:
\begin{equation}
L_{total}= \sum_{i=0}^{N+1-n_1,2} (2i+1)L_i.
\end{equation}
The quantum number $L_i$ of each block simply increases like $L_i=i$ and hence
\begin{equation}
\label{LMBn}
\langle L_{MB} \rangle_{n_1}=\frac{\sum L_i (2i+1)}{\sum (2i+1)}=\frac{N'(4N'+7)}{6(N'+1)}\simeq \frac{2}{3} (N-n_1),
\end{equation}
where $N'=N-n_1$. In the limit $N\gg1$ we get:
\begin{equation}
 \label{LMBd}
\langle L_{MB} \rangle_{d_1} \simeq \frac{2}{3} N d_1.
\end{equation}
The average over all MB $|\Psi\rangle$ states, of all $n_1$ simply gives:
\begin{equation}
\label{Lall}
 \langle L_{MB} \rangle_{all~states}=\frac{N}{3}.
\end{equation}
On the other hand, a MF state, with equidistributed excited bosons:
\begin{equation}
\label{sym_perm}
\phir=|n_1,n,n,n\rangle,
\end{equation}
with $n_1+3n=N$, has no well-defined angular momentum quantum number $L$ (except from the single case $|N,0,0,0\rangle$). We can, though, calculate the expectation value on a state $\phir$ from Eq. (\ref{L}) as:
\begin{equation}
\label{L-expect}
\langle \hat L^2\rangle=\langle \vec{n}| \hat L^2|\vec{n}\rangle=n_2(n_3+1)+n_4(n_3+1)+n_3(n_4+1)+n_3(n_2+1)+(n_2-n_4)^2
\end{equation}
for a general permanent 
\begin{equation}
\label{asym_perm}
 \phir=|n_1,n_2,n_3,n_4\rangle,
\end{equation}
or
\begin{equation}
\langle \hat L^2\rangle=4n(n+1)
\end{equation}
for the permanent of Eq. (\ref{sym_perm}). For comparison purposes, we define a \textit{pseudo-quantum number} $\tilde L_{MF}$, such that 
\begin{equation}
\tilde L_{MF} (\tilde L_{MF}+1)=\langle\hat L^2\rangle.
\end{equation}
Hence:
\begin{equation}
\label{Ltil}
\tilde L_{MF}=\frac{1}{2} \left(-1+ \sqrt{16n^2+16n+1}\right)=\frac{1}{2} \left[-1+\sqrt{\frac{16}{9}\left(N-n_1\right)^2+\frac{16}{3} \left(N-n_1\right)+1}\right].
\end{equation}
For $N\gg 1$, we get:
\begin{equation}
\label{LMFeqLMB}
 \tilde L_{MF}=\frac{2}{3}\left(N-n_1\right)\simeq\langle L_{MB} \rangle_{n_1}.
\end{equation}

So, indeed the angular momentum of the MF state (\ref{sym_perm}) equals, under the assumption $N\gg1$, the mean angular-momentum of the MB states with the same s-depletion.
Equation (\ref{LMFeqLMB}) immediately gives back Eq. (\ref{LMFdep}):
\begin{equation}
\label{LMF-dep}
\frac{\tilde{L}_{MF}}{N}=\frac{2}{3} \left(1-\frac{n_1}{N}\right)=\frac{2}{3} d_1.
\end{equation}

Now, the average angular momentum over the permanents of Eq. (\ref{asym_perm}) with the same $n_1$ is:
\begin{equation}
 \langle \tilde L_{MF}\rangle_{n_1} =-\frac{1}{2} +\sqrt{\frac{3N-8n_1}{6}N},
\end{equation}
for $N\gg1$ and $N>3n_1$, whereas the average over all the permanents of Eq. (\ref{asym_perm}) reads:
\begin{equation}
  \langle \tilde L_{MF} \rangle_{all~states} = \frac{N}{\sqrt{5}} = \frac{3}{\sqrt{5}} \langle L_{MB} \rangle_{all~states},
\end{equation}
also at $N\gg1$.

Last, we prove the condition for a MF state of Eq. (\ref{asym_perm}) to be eigenstate of the angular momentum operator $\hat L^2$, already given in Sec. \ref{MFpredictions}. Let $\phirl$ be a single-permanent eigenstate of $\hat L^2$ of Eq. (\ref{L2}). Then it must 
\begin{equation}
 \label{Leiperm}
\hat L^2 \phirl=L(L+1) \phirl,
\end{equation}
where $L(L+1)$ is the eigenvalue of $\hat L^2$ for this permanent. Then from Eq. (\ref{L}) we get that the conditions:
\begin{equation}
\begin{cases}
n_3=0   & \mbox{or} ~~ n_3=1~~  \text{and}\\
n_2=0   & \mbox{or} ~~  n_4=0 
\end{cases}
\end{equation}
must hold simultaneously. From here it turns out that the permanents that can satisfy Eq. (\ref{Leiperm}) are:
\begin{eqnarray}
\phirl=|N+M_L,0,0,-M_L\rangle, & \text{with} & M_L=-L, \\
\phirl=|N-M_L,M_L,0,0\rangle, & \text{with} & M_L=L,  \\
\phirl=|N+M_L-1,0,1,-M_L\rangle, & \text{with} & M_L=-L+1, L\geq1, \\
\phirl=|N-M_L-1,M_L,1,0\rangle, & \text{with} & M_L=L-1, L\geq1,
\end{eqnarray}
where $N$ the total number of particles and $M_L=n_2-n_4$ the quantum number of $\hat L_z$, as usual. Thus we see that the only permanents that can be eigenfunctions of the operator $\hat L^ 2$ are the permanents with quantum numbers restricted to:
\begin{equation}
 \label{MLL}
M_L=\pm L, \pm(L-1).
\end{equation}

Unless $\lo=0$, Eq. (\ref{MLL}) serves as a necessary but not sufficient condition, for a MF state to be eigenstate of both the angular momentum operators $\hat L^2, \hat L_z$ and also the Hamiltonian $\hat H$. In the case of $\lo=0$ there are no couplings among states with the same $L$ and $M_L$ and condition (\ref{MLL}), hence, suffices to determine a MF eigenstate of $\hat L^2, \hat L_z$ and $\hat H$. The same is expected to happen for small values of $\lo$.

\clearpage

\end{document}